\documentclass[12pt,a4paper]{article}
\usepackage{epsfig,graphicx}
\newcommand{\be}{\begin{equation}}
\newcommand{\ee}{\end{equation}}
\newcommand{\bea}{\begin{eqnarray}}
\newcommand{\eea}{\end{eqnarray}}
\newcommand{\beas}{\begin{eqnarray*}}
\newcommand{\eeas}{\end{eqnarray*}}
\newcommand{\ba}{\begin{array}}
\newcommand{\ea}{\end{array}}
\newcommand{\nn}{\nonumber}

\newcommand{\bt}{\begin{table}}

\newcommand{\vsi}{\varsigma}

\newcommand{\al}{\alpha}

\newcommand{\de}{\delta}
\newcommand{\De}{\Delta}
\newcommand{\ka}{\kappa}
\newcommand{\la}{\lambda}
\newcommand{\La}{\Lambda}
\newcommand{\na}{\nabla}

\newcommand{\si}{\sigma}

\newcommand{\ti}{\tilde} 

\begin{document}

\title{\bf
Unimodular bimode gravity and\\ 
the coherent scalar-graviton field\\  as
galaxy dark matter}
\author{Yu.~F.~Pirogov
\\
\small{\em Theory Division, Institute for High Energy Physics,  Protvino, 
Moscow Region, Russia }
}
\date{}
\maketitle

\begin{abstract}
\noindent
The explicit violation of  the general gauge invariance/relativity is adopted as
the origin  of  dark matter and dark energy of the gravitational  nature.
The violation of the  local scale invariance alone,  with  the 
residual unimodular one,  is considered.  
Besides the  four-volume preserving deformation mode --  the 
transverse-tensor graviton --  the metric   comprises  a compression mode
-- the scalar graviton, or  the {\em systolon}.
A  unimodular invariant and general covariant  metric theory of the 
bimode/scalar-tensor gravity 
is consistently worked out. To  reduce the  primordial ambiguity of the
theory a dynamical global symmetry is imposed, 
with its subsequent spontaneous breaking revealed.  
The  static spherically symmetric case in the empty, but
possibly for the origin, space is studied. A three-parameter solution describing
a new static space structure --  the dark  {\em  lacuna} --  is constructed.
It enjoys  the property of gravi\-tational
confine\-ment, with the logarithmic potential of gravitational attraction    at
the periphery, and results in the asymptoticaally  flat rotation curves.  
Comprising  a super-massive dark {\em fracture} (a scalar-modified black hole)
at the origin surrounded by a  cored dark {\em halo}, 
the dark lacunas are proposed  as  a prototype model  of galaxies,
implying an ultimate account for  the  distributed non-gravitational
matter and a putative asphericity or rotation.
\\[1ex]
{\bf PACS numbers:}  04.50.Kd Modified theories of gravity --   95.35.+d Dark
matter --  95.36.+x Dark energy
\end{abstract}
\vspace{1ex}

\section{~Introduction and motivations}
\label{intro}

The General Relativity (GR) is well-known to be the viable theory of
gravity perfectly consistent (modulo some additional assumptions) with all the
available observations. Nevertheless,
it hardly  is an ultimate theory  being rather  an effective field theory
produced by  a more fundamental one. 
The latter well may enjoy a wider  ``low-energy'' 
remnant, with GR being just a principle part of it. 
The basic requirements to such a GR extension may be that it should
safely retain all the well-established theoretical and observational properties
of GR,  but at the same time  should encompass some new phenomena  beyond GR 
(or v.v.).  Consider these items in more detail.

\paragraph{Gauge invariance: unimodular  vs.\  general} 

The essence of GR  may be expressed by saying  
that it is a gauge theory of
a massless tensor  field/graviton. As the respective gauge group it is  
conventionally taken the group  of the general diffeomorphisms. 
For consistency with~GR, an extension to the latter should  conceivably
be based on a gauge principle,~too. 
Following the original description of the 
irreducible (one-dimensional) unitary representations
of the (non-compact) Poincare group due to Wigner~\cite{Wigner},
it was later found by van der   Bij, van Dam  and Ng~\cite{vdB}  that the
necessary and sufficient  gauge group admitting a
massless tensor field is the group of transverse
diffeomorphisms, with the group of general diffeomorphisms  being thus
excessive to this purpose.   For this reason, a  theory of gravity based on such
a minimally violated
gauge invariance might well be   the   most natural candidate to supersede GR.
To retain  general covariance  the transverse diffeomorphisms are to be
substituted though by the unimodular ones (see Sec.~\ref{sec:2}).

To clarify the  structure of such a GR  extension  decompose  the group
of the general diffeomorphisms  into
the commuting subgroups of  the transverse diffeomorphisms  and 
that of the local scale transformations. 
By construction, the determinant of metric, $g$,   changes
only with respect to  the latter subgroup.   
Under restriction by the transverse subgroup, $g$ is  a scalar and can
be treated as an independent field variable  
to be implement in the gravity Lagrangian threefold.

(\/{\em i}\/) {\em General invariant Lagrangian }  First, one may  retain $g$ 
in the gravity Lagrangian implicitly
through metric, but made it unphysical by means of the local scale
transformations as in GR  (or in its modifications preserving general gauge
invariance).  As  a physical degree of freedom in metric there is left just the
(massless) tensor graviton.

(\/{\em ii}\/) {\em Restricted Lagrangian }   Second, one may  restrict the
Lagrangian by fixing $g$ a
priori (as it was originally proposed in other context by Anderson and
Finkelstein~\cite{Fin}), so that the respective degree of freedom in metric is
absent,  leaving only the massless tensor graviton (see also,
e.g.,~\cite{vdB}, \cite{Wil}--\cite{Smolin}).  Being a close counterpart of GR
such a theory of tensor gravity is conventionally   referred to as the
Unimodular Relativity (UR). A possible advantage of UR compared to GR  is that
the cosmological term  is no more  a Lagrangian parameter as in GR itself,
but appears as an integration constant. This may help in  solving 
the so-called naturalness problem of  cosmological term.

(\/{\em iii}\/)  {\em Extended Lagrangian }  And  v.v., one may  extend the
gravity Lagrangian   in the wake of  Buchm\"uller, Dragon and
Kreuzer~\cite{Dragon}--\cite{Kreuzer2}   by adding to
it   the terms explicitly containing    the derivatives of $g$.
Due to violation of the  general gauge invariance,  the degree of freedom
corresponding to $g$  can not  be eliminated  any more and
gets physical. Thus, in addition to the  massless  tensor  graviton the 
metric comprises a (light)  gravitationally interacting scalar  particle.  
Studying  such a  scalar-tensor  extension to GR is 
of  considerable  interest by itself from the  theoretical point of view.

\paragraph{Unimodular invariance and dark matter}

However, the main motivation to adhere to  the theory of the  latter type 
may lie in the problems of the so-called dark matter (DM) and dark
energy (DE).\footnote{For
a review on these topics  see, e.g.,~\cite{DM}.}   These substances, being
crucial for the structure of the Universe, present  perhaps  the greatest
challenge facing  the modern physics.  Basically, while DE should be
``spilled'' all over the  Universe on the cosmological
scale, DM should cluster on the  galactic and galaxy cluster  scales.   At that,
DM proves to be one of the most important
building-blocks of the Universe. Its energy yield  largely exceeds that of 
luminous  matter.
While DE might already  be anticipated as given by the well-known
cosmological term, DM appears, surprisingly enough,  absolutely ad hoc from the
modern theory point of view. Although the evidence for DM is
overwhelming, its nature remains still obscure.

The crucial points concerning DM are,  first, that it should
interact with the luminous matter very faintly and, second, its clusterization
in  gravity field should be  weaker compared to the 
luminous  matter. Such an elusive kind of matter may well  have an
unconventional nature, e.g., the  gravitational one. 
With this in mind, one can  put forward   the                               
hypothesis that  as an origin of such a DM  there  may serve
the explicit violation of the general gauge 
invariance/relativity~\cite{Pir1}.\footnote{The
term ``relativity/invariance  violation''  is  more appropriate (see
Sec.~\ref{sec:2}) 
than the ``covariance violation'' used previously in~\cite{Pir1}.}
The reason is that the ensuing in this case  extra terms in 
gravity equations may be treated as 
an effective  energy-momentum tensor of some additional
physical  degree(s) of freedom. Interacting only gravitationally, but
otherwise than  the non-gravitational matter, this new substance 
may well suit as DM.  Thus 
due to violation of general relativity the 
metric field  itself may store  a kind of the  gravitational DM.\footnote{ This
does not exclude, of course, a fraction of a more conventional particle DM, so
that the direct searches for DM are by no means meaningless.} 

As a paradigm one may consider
the minimal violation of general relativity to the  unimodular
one~\cite{Pir1,Pir1a}. In this case, on the one hand, the residual  invariance
suffices to justify the masslessness of  tensor graviton. On the other hand, due
to violation  of the local scale invariance alone,
the metric comprises  no more than one extra physical component
besides the massless  tensor graviton. 
Originating from  metric and interacting only gravitationally, the respective
particle may naturally be associated with the scalar  graviton. To have sense in
the arbitrary observer's coordinates the theory
should be  cast to the general covariant form by introducing a 
non-dynamical scalar density of the same weight as $\sqrt{-g}$. By this token,
it was found appropriate to treat  the
scalar-graviton field   as an independent variable, embedding within it
(virtually)  the  unknown scalar density. In the static spherically symmetric
case  this made it possible  to  find an exact solution to
the extended gravity equations in the empty, except for the origin,
space~\cite{Pir2}. Such a solution describes a
two-parameter generalization of the one-parameter black holes (BHs) and is valid
also in GR with  a (free massless) scalar field. 
Beyond  GR,  an approximate,  regular at the
origin  one-parameter solution  in the empty space was then  found
analytically~\cite{Pir3} and refined numerically~\cite{Pir4}. 
This solution presents a halo-type static space structure built entirely of
scalar gravitons. Such a   structure was shown to  possess a soft-core
energy density profile, qualitatively compatible with that for the galaxy DM
halos. This makes it urgent to investigate the more general solutions to the
aforementioned equations, as well as  consequences thereof.

\paragraph{Content}

In this article the   metric theory of the scalar-tensor  gravity built
on principles of the  unimodular relativity and general covariance  is
consistently worked out. In Section~2 the theoretical
background is  developed.
In Sec.~\ref{sec:2}   the explicit violation of general
relativity is  discussed in toto.   The violation of the
local scale invariance alone, with    the residual unimodular invariance, is
then investigated in more detail.   In Sec.~\ref{sec:3} 
the effective field theory of metric,  enjoying    the residual unimodular
invariance as well as    the
general covariance, is considered. To terminate
a priori allowed effective Lagrangians, a dynamical
global symmetry at the classical level is imposed. The ensuing classical
equations are then written down, and the spontaneous breaking of the 
global symmetry is displayed.  
Section~3 deals with the ensuing classical  equations and their
solutions in the static spherically symmetric case. 
In Sec.~\ref{sec:4}  the gravity equations in the empty, but
possibly for the origin, space are presented in a specific gauge. 
In Sec.~\ref{sec:5} an exact two-parameter solution of the BH-type, valid also
in  GR with a (free massless) scalar field, is rebuilt in the most transparent
fashion, and then compared in several gauges. 
In Sec.~\ref{sec:6}  an approximate,  regular at the origin one-parameter 
halo-type solution, missing in GR with a free  scalar
field, is exposed. In Sec.~\ref{sec:7} an irregular  at the origin
three-parameter  solution interpolating between  the two preceding extreme cases
is constructed. The property of the  gravitational confine\-ment for the
respective  extended static space  structures is found.   
Section~4 is devoted  to interpretation and application
of the found solutions/structures in the context of galaxy DM.
In~Sec.~\ref{sec:8} the energy content of the  structures is revealed.  
In Sec.~\ref{sec:9}  the asymptotic flat rotation curves (RCs)  ensuing due
to  these structures are exposed,  with their
DM interpretation  presented.  Finally, the relevance of the
structures to  galaxies is discussed. In Conclusion the state of affairs of the
theory and its future prospects are  outlined.

\section{~Theoretical  frameworks}

\subsection{General relativity violation}
\label{sec:2}

\paragraph{Relativity/invariance vs.\ covariance}

To go beyond GR let us refine the paradigm of  general relativity
and/or general  covariance adopted in GR. Let $I=\int{ \cal L} d^4x$ be the 
action of a field theory, with
$ {\cal L}$ being the density of  its effective  Lagrangian. To be precise,
under the (classical)  dynamics there will be  understood   the principle of
the minimal action, with the emerging classical  equations. In this respect, 
$\cal L$~may depend on two kinds of fields: the 
 {\em dynamical\/}/{\em \/relative} 
and  {\em non-dynamical\/}/{\em \/absolute} ones, generically, $\varphi$'s and
$\varphi_\ast$'s.  The   non-dynamical fields are given
a~priori. The   dynamical ones are those,  classical equations
for which are obtained  by  extremizing $I$, under frozen
$\varphi_\ast$'s.   The $\varphi$'s include
metric $g_{\mu\nu}$ (or its restricted part),  matter fields and, optionally,
the undetermined Lagrange
multiplier  which is a kind of the dynamical scalar field. The classical
solutions for $\varphi$'s should depend on $\varphi_\ast$'s  as on the external
functional parameters. An arbitrary effective  field theory, in contrast
to GR, may thus  be said to be   the theory of   ``restricted relativity'',
with $\varphi$'s implemented in the relative fashion and $\varphi_\ast$'s  in
the
absolute one.

Accordingly, there are envisaged  two types of  space-time properties of 
the field theory.

({\em i\/}) {\em Covariance}\hspace{1ex}  This  is a kinematic property  
describing a
maximal kinematically allowed set $G$ of  the simultaneous diffeomorphisms  of
$\varphi$'s  and $\varphi_\ast$'s, under which  $I$ remains  invariant. 
For consistency,  $\varphi_\ast$'s should  transform  under $G$ through
themselves, whereas the transformed $\varphi$'s may depend on  $\varphi_\ast$'s
as well. To have sense in the arbitrary observer's
coordinates (chosen a priori) $\cal L$  should be  a  general
covariant scalar density due to  a required number of $\varphi_\ast$'s (if any).
In what follows,  $G$ will be supposed to be the group of general
diffeomorphisms.

({\em ii\/})  {\em Relativity}\hspace{1ex}  This is a  dynamic
property describing  a maximal subset $H\subseteq G$ of the diffeomorphisms  
of  $\varphi$'s  alone (with $\varphi_\ast$'s transforming 
as scalars), under which $I$ remains invariant.  At that, in the process of
variation any $\delta\varphi$ within the  configuration space are admitted,
whereas $\varphi_\ast$'s  are to be  frozen ($\de\varphi_\ast=0$). 
By the very nature,
it is $H$ which serves as a space-time gauge group for the theory. 
So, as a synonym of relativity  there will also be used the term (gauge) 
{\em invariance}. 

In GR  the covariance and  
relativity/invariance coincide, both being the general ones. Beyond GR 
one should   distinguish them. In particular,  the non-trivial quotient $G/H$
may result in appearance of the extra physical degree(s) of freedom.
In what follows  we consider the unimodular invariant but general covariant 
metric theory of gravity,   with a  dynamical
scalar-graviton  field in metric and  a non-dynamical scalar-density field.
This is the maximal allowed explicit violation of general
relativity/invari\-ance consistent with  the
masslessness    of  
transverse-tensor graviton.  

\paragraph{Unimodularity and bimodality}

To be precise, under the theory of gravity  at the energies lower than the Plank
mass  there will be  understood  the effective field theory of metric.
Ultimately, the latter theory  is to be considered as a ``low-energy''
remnant   of a more fundamental theory.  This remnant may   basically be
characterized  by the  two ingredients: a
residual invariance group~$H\subseteq G$   and  a set of the low-energy fields.
As a paradigm,  let $H$ be the  maximal subgroup of
$G$ given by the unimodular group $U$. 
Let $g_{\mu\nu}$ be a dynamical metric field, with ${\rm det}(g_{\mu\nu})=g$,
and let $\sqrt{-g_\ast}$ be  a non-dynamical scalar density of the same weight
as  $\sqrt{-g}$. Related to DM and associated with a non-dynamical measure, 
$\sqrt{-g_\ast}$ will be, 
in the wake of~\cite{Fin},   referred to  as the   (dark) {\em    modulus}. 
A priori  there are envisaged arbitrary (including singular) moduli. The
dark moduli  are to be considered as  a kind of a  new substance, the nature of 
which in the framework of the effective field theory is unspecified  and  should
ultimately  
be revealed by a more fundamental theory.

Without loss of generality,  the Lagrangian density ${\cal L}$ for a
unimodular invariant effective field theory of metric may be expressed
through the general covariant  scalar Lagrangian~$L$ as 
${\cal L}= L(g_{\mu\nu}, g,  g_\ast)\sqrt{-g}$.
The gauge properties of a field theory under the
infinitesimal diffeomorphisms $x^\mu\to \hat x^\mu= x^\mu +\De_{\xi}  x^\mu$,  
$\De_{\xi}  x^\mu  =-\xi^\mu$, are expressed through  a Lie derivative. The
latter accounts for the net variation of a field due to both its  tensor and
argument variations. For the metric field  it is as follows:
\bea
\De_{\xi}g_{\mu\nu}& =&  \partial_\mu \xi^\la g_{\la\nu}+
 \partial_\nu \xi^\la g_{\mu\la}
+\xi^\la\partial_\la g_{\mu\nu}=\nn\\
&=&\nabla_{\mu}  \xi_{\nu}+\nabla_{\nu} \xi_{ \mu} ,
\eea
with $\xi_\mu = g_{\mu\la}\xi^\la$ and $\nabla_\mu$ being  a covariant
derivative  with respect to  $g_{\mu\nu}$.  It follows hereof that 
\be
\De_{\xi}\sqrt{-g}  =\frac{1}{2}\sqrt{-g}g^{\mu\nu}\De_{\xi}g_{\mu\nu}  =
\partial_\mu( \sqrt{-g} \xi^\mu ).
\ee
By definition, the same should fulfill for $\De_{\xi}\sqrt{-g_\ast}$.
Thus
\be
\De_{\xi}\ln \sqrt{g/g_\ast}= \xi^\mu \partial_\mu\ln \sqrt{g/g_\ast},
\ee
i.e.,  $g/g_\ast$ transforms under $G$ as a scalar.

Now, restrict  the set of $\xi^\mu$ by a subset  $\xi_u^\mu$ defined by  
requirement of invariance of  a dark  modulus (the {\em unimodularity}
condition):\footnote{Restricted ab
initio to coordinates where $g_\ast=-1$ (if possible),
the covariant unimodularity condition  reduces to the non-covariant
transversality condition: $\partial_\mu
\xi^\mu_t=0$, with $\xi^\mu_t=\xi^\mu_u|_{g_\ast=-1}$. For the 
``transverse/TDiff gravity'' based on the latter condition  see, e.g.,
\cite{Alvarez}--\cite{Shap2}.}
\be
\De_{\xi_u}\sqrt{-g_\ast}  =\partial_\mu( \sqrt{-g_\ast} \xi_u^\mu )=0.
\ee
This  singles out in a  covariant fashion  the residual unimodular  
invariance subgroup  $H=U$.  At that 
\be
 \De_{\xi_u}\ln\sqrt{-g}  =   \xi_u^\mu \partial_\mu\ln\sqrt{g/g_\ast}
\ee
is not bound to  vanish.
The general covariance forbids  dependence of $L$ on $g_\ast$ and $g$
separately,  leaving just 
\be\label{UM}
{\cal L}=  L(g_{\mu\nu}, g/g_\ast)\sqrt{-g}.
\ee

An arbitrary $g_\ast$
is unavoidable in the general covariant framework of the unimodular invariant
field theory, despite the fact that it might  superficially seem  irrelevant.
Besides,
it is urgent   in practice.  In addition to the implicit manifestations it may
conceivably  result in some direct ones. The  terms  violating general
invariance and containing  derivatives of
$g/g_\ast$ result ultimately in  the scalar 
propagating mode in metric.  
In the absence of derivatives,
$g/g_\ast$ becomes  just an auxiliary field.
In what follows the general covariant  and unimodular invariant
metric theory of the bimode/scalar-tensor gravity -- the  Unimodular Bimodal
Gravity (UBG) -- is consistently exposed.\footnote{To be 
distinguished from the ``plain''   (by default, mono\-mode/tensor)  UG
sometimes used as  a synonym of UR.   Under the  phenomenon of unimodular
gravity  we understand generically both the
unimodular bimode and 
monomode/tensor gravities, the latter  of them treated as a marginal case. The
term bimode gravity refines the more sophisticated  one --  ``metagravity'' -- 
used previously in~\cite{Pir1,Pir1a}.}

\subsection{Unimodular Bimodal Gravity}
\label{sec:3}

\paragraph{Extended gravity Lagrangian}

In the framework of the effective field theory
any choice ${X}={X}(g/g_\ast)$  is a priori
allowed as a field variable instead of $g/g_\ast$. The different choices are 
related through  the field redefinition.  
Thus let us  take without loss of generality as the
(dimensionless) scalar-graviton  field 
\be\label{Sg}
{X}=\ln \sqrt{-g}/ \sqrt{-g_\ast}.
\ee
This  choice  proves to be  advantageous from the symmetry considerations (see
below). In this terms,  one has   for  the  UBG Lagrangian generically
\be
L_{\rm \small UBG}=L_g+ L_s  + \Delta L_{gs} + L_m,
\ee
where
\begin{equation}\label{LGS}
L_g=- \frac{1}{2}  \kappa_g^2R, \ \ 
L_s =  \frac{1}{2}  \kappa_s^2\na {X}\cdot  \na {X} 
\end{equation}
present  the  kinetic terms, respectively, for the  tensor and scalar modes,
 with  $\na {X}\cdot\na {X} =g^{\mu\nu} \na_\mu {X} \na_\nu
{X}$, etc., and $\na_\mu X\equiv \partial_\mu X$.
The general  invariant $L_g$ is chosen for simplicity  as in GR (cf.\ though
Sec.~\ref{sec:9}). 
Here  $ R$ is  the Ricci scalar and $\kappa_g=1/\sqrt{8\pi
G}$ is the  mass scale for tensor
gravity, with $G$ being the Newton's constant.
In the   unimodular invariant $L_s$,  the parameter   $\kappa_s$ is
a mass scale for the scalar gravity. 
For  tensor dominance of the bimode gravity  (generically
$|L_g|> |L_s|$) one would a priori  expect $\ka_s/\ka_g< {\cal O}( 1)$. 
Moreover, to comply with  astronomic observations for galaxies
 it proves that $\kappa_s/\kappa_g \sim 10^{-3}$ (see Sec.~\ref{sec:9}).  
Further, $\Delta L_{gs}(R, {X}, \na_\mu {X},\dots)$ is the
rest of  the gravity Lagrangian depending, generally,   on  $R$  and $ {X}$.
 In particular, $\Delta L_{gs}$ may include  some  scalar potential, 
$\Delta L_{gs} =-V_s({X})+\dots$.  At that, the  non-vanishing asymptotically
constant  part of the potential, $V_s|_\infty$,  
may be attributed to  the cosmological term. 
The Lagrangian for gravity is to be  supplemented  
by that for  the non-gravitational matter, $L_m(\psi, R, {X},\na_\mu {X},
\dots)$, with $\psi$ designating a generic matter field. In principle  the
latter  may correspond both to the luminous matter and a
putative non-gravitational  DM.
Finally, instead of being imposed
explicitly,  the restriction~(\ref{Sg}) may be
enforced classically with the help of the
undetermined Lagrange multiplier $\la$ by adding to $L$  the term
\begin{equation}\label{Lla}
L_\la=\la\Big( \sqrt{- g_\ast} /  \sqrt{- g}  -    e^{-{X}} \Big).
\end{equation}
This allows one to  treat ${X}$ as  an independent variable.

\paragraph{Global compression  symmetry}

First of all, in the spirit of the
effective field theory, the extra terms in Lagrangian containing  
derivatives of ${X}$  are  to be suppressed  compared to  the kinetic term for
${X}$
by powers of $1/\kappa_s$, and thus may  be  neglected.
To further terminate the Lagrangian,  enhance the residual unimodular 
invariance  by a dynamical global symmetry defined in the fixed coordinates
through  the field substitutions  as follows:
\bea\label{gt}
g_{\mu\nu}(x)& \to& \hat g_{\mu\nu}(x)=k_0^2 
g_{\mu\nu}  (k_0 x),\nn\\
g(x)&\to & \hat g(x) = k_0^8 g(k_0x )   ,\nn \\ 
g_\ast(x)&\to   &  \hat g_\ast(x) = 
g_\ast (k_0x )   ,
\eea
with $k_0>0$ being an arbitrary  constant.
This is a generic  symmetry distinguishing the  dynamical  and non-dynami\-cal 
fields. For the former ones  it coincides (modulo the coordinate redefinition) 
with the conventional global scale symmetry, being a part of the
general coordinate transformations. For this reason, the general invariant part
of $L_{\rm\small  UBG}$ ($L_g$ and, supposedly, $L_m$) is
global symmetric.  A  common multiplicative factor appearing in $\cal
L$ due to $\sqrt{-g}$ does not influence the classical equations.
Eq.~(\ref{gt})
will be referred to as the {\em compression} transformations. In these terms,
the  moduli are ``incompressible'' in contrast to the dynamical metric. 
Further, it follows from (\ref{gt}) that   ${X}$ transforms inhomogeneously
under the compressions: 
\be\label{gt{X}}
{X}(x)\to  \hat {X} (x)= {X}(k_0  x) +4 \ln k_0.
\ee
The emergence of the respective  (approximate) global  symmetry  may
serve as a natural reason for suppression of   the derivativeless   couplings of
the scalar graviton  as   vio\-lating  the global  symmetry (e.g., $V_s({X})$,
etc.). Due to~(\ref{gt{X}}) the scalar graviton may be
treated as a (pseudo-)Goldstone boson for a hidden/non-linear  realization of
the  compression symmetry. Associated with the scale invariance,
such a scalar graviton/(pseudo-)Goldstone boson may
more spe\-cifically be termed the {\em systolon},\footnote{From the Greek
$\sigma\upsilon\sigma\tau o \lambda\eta\! \!'$ meaning the compression,
contraction.} with  the tensor
graviton being conventionally  just the graviton. 
The systolon presents a scalar compression mode in metric in addition to the
transverse-tensor, four-volume preserving deformation  mo\-de  presented by the
(massless) graviton. Finally, imposing  
\be\label{gsla}
\la(x)\to  \hat \la(x) =k_0^{4} \la(k_0 x), 
\ee
one  gets that $ L_\la$ does not violate  global  symmetry, too.

The (approximate) global compression symmetry may be used as 
the third  basic ingredient of UBG, in
addition to  general
covariance and unimodular invariance/relativity.
To such a reduced version of  the  theory   we will adhere in what follows. This
is a minimal bimode extension to UR. Namely, from (\ref{LGS}) and (\ref{Lla}) in
a formal limit of ``switching-off''  the scalar mode, ${X}\to  0$,  one recovers
 at the classical level  UR, the latter, in
turn, being
classically equivalent to GR with a cosmological term. 
For more gene\-rality,  we still  retain  the   terms
$V_s({X})$  and $L_m({X})$ assuming  that they are  the leading
corrections to the otherwise  global  symmetric Lagrangian.\footnote{UBG  may be
considered as a general covariant counterpart of a theory of
gravity~\cite{Dragon}--\cite{Kreuzer2} comprising in metric  the (tensor)
graviton and dilaton,  but restricted  ab initio (hereof the term ``restricted
gravity'') exclusively to coordinates  corresponding in UBG to 
$g_\ast=-1$. The general covariance proves though to be
crucial for real treating the theory   and  associating it with
DM. In passing, it is the gauge invariance which is in fact restricted, the
gravity itself  being rather extended. }

\paragraph{Extended gravity equations}

Varying the total action (assuming $\Delta L_{gs}= - V_s$) with
respect to
$g^{\mu\nu} $ and ${X}$ independently one gets  the  tensor  and scalar gravity 
equations, respectively,  as follows:
\begin{eqnarray}\label{mgeq}
  \kappa_g^2\Big(R_{\mu\nu}-\frac{1}{2}R g_{\mu\nu}\Big)-\la 
e^{- {X}} g_{\mu\nu} &=&T_{m \mu\nu}  +t_{s \mu\nu}  ,\nn\\
\kappa_s^2\nabla\cdot\nabla  {X}+  \partial V_{s}/\partial {X}-\la
e^{-{X}}&=&\partial L_m/\partial {X},
\end{eqnarray}
with 
\begin{equation}\label{d'Alambert}
\nabla\cdot\nabla  {{X}}
= g^{\mu\nu}\nabla_\mu\nabla_\nu  {{X}}
=\partial_\mu(\sqrt{-g} g^{\mu\nu} \partial_\nu  {{X}})/\sqrt{-g}
\end{equation}
being  the general covariant d'Alambertian operator. Conventionally, 
\be\label{T_m}
T_{m \mu\nu}=\frac{2}{\sqrt{-g}}\frac{\partial (\sqrt{-g}  L_m)}{\partial
g^{\mu\nu}}=2 \partial L_m/\partial g^{\mu\nu}-  L_m   g_{\mu\nu}
\ee
is the canonical energy-mo\-men\-tum tensor of the non-gravi\-tational 
matter, and
$t_{s\mu\nu}$ is the similar  scalar-field  tensor  given
by the unconstrained~$L_s$: 
\begin{equation}\label{enmom}
t_{s\mu\nu}=\na_\mu  \si\na_\nu  \si
-\Big( \frac{1}{2} \na \si\cdot\na \si -V_{s}
\Big)g_{\mu\nu} .
\end{equation}
In the above  we have conventionally introduced  the dimensionfull scalar field
$\si=\ka_s
{X}$.
Introducing $W_s$ as $\kappa_s$ times  the canonical inhomogeneous scalar-field 
wave operator:
\begin{equation}\label{W}
W_{s}= \kappa_s(\nabla\cdot\nabla  \si+  \partial V_{s}/\partial \si-\partial
L_m/\partial \si)
\end{equation}
one may present   the scalar gravity equation as follows
\be\label{Wla}
 W_s= \la e^{-{\si/\kappa_s}},
\ee 
with the r.h.s.\  playing the role of a (virtual) source density. 

Excluding $\la$ from (\ref{mgeq})    we finally arrive at the extended 
gravity equations in the superficially  conventional form
\begin{equation}\label{eg}
R_{\mu\nu}-\frac{1}{2}Rg_{\mu\nu} = \frac{1}{\kappa_g^2}{T}_{ \mu\nu}, \  \
{T}_{\mu\nu}= T_{m \mu\nu}+ { T}_{s \mu\nu}.
\end{equation}
Here ${T}_{ \mu\nu}$ is  the total   ``bare'' 
energy-momen\-tum tensor, with the tensor gravity included only in the minimal
fashion through  metric.  At that, the bare effective    tensor 
for systolons 
\be\label{Ts}
{ T}_{s \mu\nu }= t_{s \mu\nu }+W_s g_{\mu\nu}
\ee  
looks like the canonical one,  $t_{s \mu\nu }$, except for the Lagrangian
potential $V_{s }$  substituted by the  effective one
\be\label{UVW}
{\La}_s= V_{s}+W_s, 
\ee
with $W_s$ given by (\ref{W}).
Otherwise, $ { T}_{s \mu\nu }$  may formally be brought to the form: 
\begin{equation}
{ T}_{s \mu\nu}=( \rho_s+ p_s) n_\mu n_\nu  +
\rho_s g_{\mu\nu},
\end{equation}
where $n_\mu\equiv\nabla_\mu \si/(-\nabla \si\cdot \nabla \si)^{1/2}$ ($n\cdot
n=-1$) and
\be
 \rho_s = -\nabla \si\cdot \nabla \si/2  + \La_s,\ \    p_s =
-\nabla \si\cdot \nabla \si/2  - \La_s.
\ee
Superficially,  this  has little to
do with a continuous medi\-um, the   account for  the proper gravity
contribution being to this end in order (see Sec.~\ref{sec:8}).

Finally, varying $L_\la$ with respect to $ \la$ one recovers  the constraint
(\ref{Sg}) which is to be understood in (\ref{eg}). The non-canonical
contribution to ${T}_{s\mu\nu}$ appears ultimately from the scalar-field kinetic
term due to such a constraint,  missing in GR with a scalar field. This may  be
inferred directly without Lagrange
multiplier, with $ {T}_{s\mu\nu}$ appearing   as a canonical
energy-momentum
tensor  from  the Lagrangian (\ref{LGS}) with the explicit 
constraint~(\ref{Sg})~\cite{Pir1}. 

Presenting  (\ref{eg})  in  the equivalent form
\begin{equation}\label{mgeq'}
R_{\mu\nu}= \frac{1}{\kappa_g^2} \Big({T}_{\mu\nu}-\frac{1}{2}{T}
g_{\mu\nu}\Big), 
\end{equation}
${T} \equiv {T} ^\la_\la={T} ^0_0+{T} ^l_l$,   $l=1,2,3$, 
we find in particular  that
\begin{equation}\label{R00}
R^0_0=  ({T}^0_0- {T}^l_l)/(2\kappa_g^2). 
\end{equation}
This is of special importance for revealing the
energy content  of a static space  structures (see Sec.~\ref{sec:8}). 
In the empty space, with $L_m=0$ except possibly for a point,
Eq.~(\ref{mgeq'}) reads
\be\label{FE}
R_{\mu\nu}= \frac{1}{\kappa_g^2}\Big (\na_\mu \si\na_\nu\si-
{\La}_s
g_{\mu\nu}\Big),
\ee
where ${\La}_s$ is given in general  by (\ref{UVW}) and (\ref{W}). 
Studying this equation and consequences thereof  in the static
spherically symmetric case is the main concern of the following. 
Under the quasi-harmonicity condition, $W_s=0$ 
except possibly for a point,  Eq.~(\ref{FE}) reduces to   the
conventional  Einstein-Klein-Gordon system in GR for a self-interacting scalar
field minimally coupled to gravity, the  solutions for  both theories thus
coinciding   in this case.

\paragraph{Spontaneous  global symmetry breaking }

Due to the  contracted Bianchi identity, 
$\nabla_\mu (R^\mu_\nu-     R/2\,   \de^\mu_\nu )=0$,
the energy-momentum tensor of the non-gravitational matter and  
systolons  collectively, not generally by parts,  is bound to
covariantly conserve,  $\nabla_\mu T^\mu_\nu=0$.
In view of  (\ref{W}) this results in the  third-order differential
equation 
\begin{equation}\label{3order}
\partial_\nu W_s + W_s \partial_\nu {\si} /\kappa_s= 
-\nabla_\mu T_m{}^\mu_\nu ,
\end{equation}
which as it stands   is hardly of use. However, in the case if $L_m$ is
independent of $\si$  (in particular,  in  empty space, $L_m=0$), the  emerging
general invariance of the matter action   implies that  
 $\nabla_\mu
T_m{}^\mu_\nu=0$. Thus $\partial_\nu(\ln |W_s|+\si/\ka_s)=0 $, and    there
appears the first integral of motion
\begin{equation} \label{echi}
W_{s} =\La_{0} e^{- \si/\kappa_s}, 
\end{equation}
with  $ \La_{0} $ being  an integration constant. It follows then from 
(\ref{Wla})  that now $\la=\La_0$.   With account for (\ref{gsla}) this
signifies
spontaneous breaking of  the global symmetry   $\si\to \si+\si_0$, with $\si_0$
being a constant. In view of  (\ref{W}), Eq.~(\ref{echi}) becomes
nothing but the bona fide  second-order  scalar-field equation:
\begin{equation}\label{subst'} 
 \nabla\cdot\nabla  {\si}+\partial {\La}_s/\partial {\si}  =0,
\end{equation}
with  the effective potential looking like  the conventional one  plus
 the exponential contribution
\be\label{Pi}
\La_s = V_s + \La_{0} e^{-{\si}/\kappa_s}.
\ee
In this case UBG  is classically equivalent to GR with a
scalar field  supplemented by   the exponential potential   in 
Lagrangian (with an  alien for GR scale $\ka_s$).\footnote{This
statement
resembles   a similar one  about the classical equivalence between  UR
and GR supplemented by  the  cosmological  term in Lagrangian. This 
can be seen   at $V_s=0$ in a  formal limit
$\si\to 0$.}
At $\La_{0}=0$ the theories moreover 
classically coincide. For physical reason, there should fulfil
$\La_{0}\leq 0$  (see Sec.~\ref{sec:8}). This  means that due to a  potential
well produced by a systolon condensate,   the  local vacuum with  the
spontaneously broken global symmetry lies lower than the  symmetric
one~($\La_{0}=0$), ensuring ultimately  spontaneous breaking of the symmetry.

\paragraph{UBG and beyond}

Let us make  some remarks posing  
UBG among  the  related  metric 
theories of gravity with  the additional degree(s) of freedom, which may have
bearing to~DM.

(\/{\em i}\/) {\em Unimodular scalar-tensor gravity}\hspace{1ex}  
First of all, put\-ting  $ g_{ \mu\nu} \equiv
(g/g_\ast)^{1/4} g_{u \mu\nu}$, with $  \mbox{\rm det\,} 
g_{u \mu\nu} \equiv g_u=g_\ast$, one  may equivalently present~(\ref{UM}) as
${\cal L}=L( g_{u  \mu\nu},  {X} )\sqrt{-g_\ast}$,
where ${X}= {X}(g/g_\ast)$ may be chosen, e.g., as in
(\ref{Sg}).
Under the global
compression symmetry  one has $g_{u\mu\nu}(x)\to
\hat g_{u\mu\nu}(x)=g_{u\mu\nu}(k_0 x)$, while
$g_\ast$ and ${X}$ transform as before. 
Now  choosing
$\si\equiv \kappa_s {X}$ as an independent scalar field one may  reduce the
original theory to the 10-field unimodular scalar-tensor
gravity,  which proves to be  nothing but the unimodular bimode gravity in
disguise. The inverse transition is achieved  by putting $ g_{ u\mu\nu} \equiv
e^{-\si/2\ka_s} g_{\mu\nu}$ and, respectively,  $\si/\ka_s =(1/2)\ln g/g_\ast$.
The two theories may thus be treated just as the generic UBG  in the
two different  frames: the bimode and scalar-tensor ones.\footnote{In the
latter frame  the unimodularity  literally
means the invariance of $g_u$.}

(\/{\em ii}\/) {\em General scalar-tensor gravity}\hspace{1ex}  
More generally, one may confront UBG, containing  10 
dynamical fields, with
the  general invariant theory  defined
by the arbitrary ${\cal L}=L(g_{\mu\nu}, \si)\sqrt{-g}$  containing
additionally an independent sca\-lar field $\si$,  altogether 11 dynamical
fields.\footnote{For a confrontation between 
 the transverse/TDiff gravity and GR with a scalar field
see~\cite{Villarejo}.}$^,$\footnote{As a
particular implementation of such a  theory there may be mentioned the
well-known Brans-Dicke model~\cite{JBD}. In the Einstein frame the
scalar field in this version of the theory acquires the  direct derivativeless
coupling with matter being thus strongly restricted observationally 
(see, e.g.,~\cite{Will}).}
As it was stated  earlier, under  the separate conservation of the
ener\-gy-momentum tensor of the non-gravitational matter  (in particular,
in the matterless vacuum) 
UBG is  classically  equivalent to
the general scalar-tensor gravity, with a  supplementary exponential
contribution to the Lagrangian potential of the latter emulating  the proper 
spontaneously emerging 
term in the classical equations of the former.
Nevertheless, in the context of DM there are  two important differences.
First,  to  match with astronomic observations the
sign of the extra contribution  is to be negative (see Sec.~\ref{sec:9}). 
Being harmless in the classical equations
such an unbounded from below term in  the Lagrangian  could
result in a quantum inconsistency.
Second, appearing as  an integration constant in the classical equations,
$\La_0$ may
vary for different solutions, whereas in  the Lagrangian potential  it should be
fixed ad
hoc ones forever.  Thus in the DM context  UBG seems   to be more safe  and
flexible compared to the general scalar-tensor gravity.

(\/{\em iii}\/) {\em Bimetric multimode gravity}\hspace{1ex}     
In a wider perspective of the general relativity violation one may consider  
the multimode gravity, with the general  covariant
Lagrangian density ${\cal L}= L (g_{\mu\nu},g_{\ast\mu\nu}) \sqrt{-g}$
depending   on the  dynamical and non-dyna\-mical metrics  $g_{\mu\nu}$ and
$g_{\ast\mu\nu}$,  respectively, as well as their determinants. 
Basically, such a theory is
characterized  by its residual gauge invariance $H \subseteq G$.  Containing 10 
independent dynamical fields,
the theory with the trivial $H=I$ may thus encounter up to 10 physical 
degrees of freedom 
(including, in general,  the ghost ones)  and present  potentially  a lot of
problems and ambiguities. To eliminate/reduce them  the
residual unimodular invariance, $H=U$,  may, e.g.,   be imposed. In
conjunction with general covariance  this  would retain  the dependence of $L$
just on $g_{\mu\nu}$ and 
$g/g_\ast$, with $ g_\ast =\mbox{\rm det}(g_{\ast\mu\nu})$, and no more than
a single extra graviton.  Nevertheless, in the wake of the developed approach to
DM one could envisage  in this direction   a multimode gravitational
DM.\footnote{For  the bimetric gravity  in  the context of the massive
tensor graviton,  with   a
priori unspeci\-fied   $g_{\ast\mu\nu}$ and extra terms only in the Lagrangian
potential,   cf., e.g.,~\cite{Visser,mg}.}

\section{~Equations and  solutions}

\subsection{Static spherical symmetry}
\label{sec:4}

\paragraph{Quasi-Galilean coordinates}

Consider the static spherically symmetric field configurations. Start with the
quasi-Galilean coordinates  $x^\mu= (x^0, x^m)\equiv(t, {\bf x}) $ ($m=1,2,3$),
where the  general covariant line element is as follows:
\begin{equation}\label{ds2}
d s^2= ad t^2 - (b- c)({\bf  n} d {\bf x})^2 - cd{\bf x}^2. 
\end{equation}
Here $\bf n$ ($ {\bf n}^2=1) $ is given by  $  n^m= x^m/ |{\bf x}| $,
with $|{\bf x}|^2={\bf x}^2 \equiv \delta_{mn} x^m x^n $.
The metric variables $a$, $b$ and $c$ are arbitrary functions of  $|{\bf x}|$
alone.  The same concerns  $ g_\ast $ (and thus~$\si$).
The respective metric looks like
\begin{equation}\label{metric}
g_{00}= a,\ \
g_{mn}=- b \bar n_m  \bar n_n
- c(\de_{mn}- \bar n_m \bar n_n) .
\end{equation}
where for short we designated $\bar n_m\equiv \de_{mn}n^n$
(vs.\ conventional $n_m=g_{mn}n^n$). The rest of the metric
elements is zero. Respectively,  the  inverse metric is
\begin{equation}\label{metric-1}
g^{00}=\frac{1}{ a},\ \ 
g^{mn}=-\frac{1}{ b}\, n^m n^n 
-\frac{1}{ c}(\de^{mn}- n^m  n^n).
\end{equation}
Due to the rotation invariance one can choose
the spatial coordinates in a point $\bf x$ so that 
$\bar{\bf n}=(1,0,0)$, bringing 
the metric in this point to the diagonal
form $g_{\mu\nu}=\mbox{\rm diag\,} (a,- b ,-c,-c)$. 
Thus     $\sqrt{-g}=\sqrt{ab}c$, and the scalar field looks like
\begin{equation}\label{sigmar}
 \si=\ka_s \ln (\sqrt{ab}c/\sqrt{-g_\ast}).
\end{equation}
The quasi-Galilean coordinates are appropriate for revealing
the energy content of the static space  structures (see Sec.~\ref{sec:8}).

\paragraph{Polar coordinates}

For real dealing with the spherical symmetry  better suit  the polar coordinates
$ x^{\mu}=  ( x^0, x^{ m})= (t, r, \theta, \varphi)$, $ m=r,
\theta,\varphi$ (with a unit of length $l_0$ tacitly understood where it is
necessary).  Conventionally, 
${\bf x}=r{\bf n}=  r(\sin\theta\cos\varphi,
\sin\theta\sin\varphi, \cos\theta$). 
The  line element now reads
\begin{equation}\label{polc} 
d s^2= a d  t^2- b d   r^2-  c  r^2 d\Omega^2, \ \
d\Omega^2 = d \theta^2+\sin^2\theta d\varphi^2,
\end{equation}
with the metric being $g_{\mu\nu}=\mbox{\rm diag} (a,-b,-c r^2,
-c r^2 \sin^2\theta)$.
The  scalar field looks as before,
with the transformation Jacobian cancelled out in the ratio of the two scalar
densities of the same weight. The functions  $a$,
$b$, $c$ and $g_\ast$ depend on~$r$ alone. 

The  systolon bare effective 
energy-momentum tensor  (\ref{enmom}), (\ref {Ts}) and (\ref{UVW}) is then
${ T}_{s}{}^\mu_\nu
\equiv \mbox{diag}(p_0,
-p_r,- p_\theta,- p_\varphi)$, where 
\begin{eqnarray}\label{rhop}
p_0=- p_\theta =- p_\varphi
&=&\frac{1}{2b}  {\si}'^2+{\La}_s ,\nn\\
p_r&=&   \frac{1}{2b}  {\si}'^2-{\La}_s,
\end{eqnarray}
with  $p_0$ being the 
bare energy density, $ p_n$ ($n=r, \theta,\varphi$) the  bare
pressure and prime
indicating  a radial derivative.\footnote{Here and in what
follows,   we define  for simplicity the energy densities  without
$\sqrt{-g}$.}
 There  follows hereof, in particular, an important  spatial invariant (see
Sec.~\ref{sec:8}):
\begin{equation}\label{trace} 
 { T}_{s}{}^{0}_{0} -{ T}_{s}{}^{n}_{ n }=  p_0+\Sigma 
p_n= -2{\La}_s.
\end{equation}

With $R^{\mu}_ {\nu}$ being diagonal and
$R^\varphi_\varphi=
R^\theta_\theta$, there is left three independent gravity equations  which may
be brought to the form:
\begin{eqnarray}\label{mgeq''}
R_0^0  &=& \frac{1}{2\ka_g^2 } \Big( T_m{}^0_0 -
T_m{}^ {n}_{n} - 2{\La}_s \Big), \nn\\ 
R_0^0- R_r^r &=&\frac{1}{\ka_g^2 } \Big (T_m{}^0_0 - T_m{}^r_r+
\frac{1}{b} {\si}'^2\Big), \nn\\ 
R_0^0  - R_\theta^  \theta   &=& \frac{1}{\ka_g^2 } 
\Big(T_m{}^0_0-T_m{}^\theta_\theta\Big).
\end{eqnarray}
In the above  the curvature elements are expressed through $a$, $b$
and~$c$. The contracted Bianchi identity reads
\begin{equation}\label{WT}
 W_s' +  W_s {\si}' /\ka_s = - T'_m{}^{r}_r.   
\end{equation}
Under the general   invariance ($\si$-independence) of $L_m$  (in particular,
at  $L_m=0$) one has $T'_m{}^{r}_r=0$, and there
appears the first  integral of motion
$W_{s}= \La_{0} e^{-\si/\ka_s} $, with $\La_{0}$ being an integration
constant.  The third-order equation~(\ref{WT}) reduces in this case to
the second-order scalar-field equation (\ref{subst'}) which looks now like
\begin{eqnarray} \label{echi'}
\nabla\cdot\nabla \si& = &-\Big(\sqrt{a/b}c r^2  {\si}'\Big)'\Big /
\sqrt{ab}cr^2 =\nn\\
&= &(\La_{0}/\ka_s) e^{-\si/{\ka_s}}-\partial V_s/\partial  {\si}  . 
\end{eqnarray}

\paragraph{Radial coordinate/gauge fixing}

The choice of  the radial coordinate $r$ is not unique. Under the local 
radial rescaling $r\to \hat r =\hat r(  r)$, the  
general covariant line element  is  moreover 
formal invariant:
$d s^2=d \hat s^2= \hat a d  t^2- \hat b d  \hat r^2- \hat c \hat
r^2 d\Omega^2$, with the relation 
\begin{eqnarray}\label{coordchange_g}
a(r)&=&\hat a(\hat r( r)) ,\nn\\ 
b(r)&=& (d \hat r/d  r)^2 \hat  b(\hat r( r))  ,\nn\\ 
c(r)  &=&  (\hat r/  r)^2 \hat c(\hat r( r)).
\end{eqnarray}
The quasi-Galilean $\sqrt{-g} =\sqrt{ab}c $ transforms under rescaling as 
\begin{equation}\label{g(r)}
\sqrt{- g(r)}= \sqrt{-\hat g(\hat r(r))}\,  
 \frac{\hat r^2}{r^2}\frac{d\hat r}{d r}
\end{equation}
(and so does $\sqrt{- g_\ast }$), with  $\si$ transforming as a scalar, 
$ \si(r)=\hat \si(\hat r(r))$.
The  three-volume element $
\sqrt{-g} d^3 x= \sqrt{ab}cr^2 d r d \Omega^2 $, being (spatial) invariant, is 
also formal inva\-riant.
It follows  from (\ref{coordchange_g})  that  metric, asymptotic
Minkow\-skian with respect to~$r$, retains this property with respect  to
$\hat
r$  if  there fulfills asymptotically  $\hat r=r(1+{\cal O}(1/r))$.
Thus, the ansatz (\ref{polc}) does not allow to fix the metric uniquely.

The aforesaid ambiguity can be removed by  the radial coordinate/gauge fixing. 
Namely, the three gravity   equations (\ref{mgeq''}) contain superficially four
variables: $a$, $b$, $c$ and $ \si$, only three of them being independent due to
constraint (\ref{sigmar}). 
To account for the latter,   two opposite routes of dealing
with  a priori unknown $g_\ast$  are envisaged.

(\/{\em i}\/) {\em Explicit $g_\ast$}\hspace{1ex} The direct route is to
assume $g_\ast$ in some starting
coordinates  and solve the equations with the proper  boundary conditions. 
Then postulating a relation between the starting and observer's coordinates, 
transform the solution to the latter
coordinates, including   boundary conditions.  In practice, one may start
from the transverse coordinates
defined by  $g_\ast=-1$,
what is equivalent to imposing the scalar-field dependent gauge
$\sqrt{-g}=e^{X}$. In the arbitrary observer's 
coordinates the solution will  thus explicitly comprise~$g_\ast\neq -1$.

(\/{\em ii}\/)  {\em Implicit $g_\ast$}\hspace{1ex}    The inverse route is to
get  (virtually)  rid  of $g_\ast$ in favour
of  $\si$.  Imposing then a   gauge
$F(a,b,c)=0$ as a definition of observer's coordinates,   solve the equations,
with the proper boundary
conditions imposed already in the latter coordinates. The dependence 
on $g_\ast$ remains implicit. Having found the solution  one can extract  hereof
 the required  $g_\ast$ solving, in a sense, an
inverse problem. Inserted back into
the  gravity equations  this  $g_\ast$ is
bound to produce in the direct route in the same observer's coordinates
precisely the given solution. Having  a priori no knowledge  about $g_\ast$
we  will  adopt in what follows such an inverse route, with several
particular gauges compared (see Sec.~\ref{sec:5}).

\paragraph{Reciprocal gauge}

It proves to be  convenient to impose the  gauge $ab=1$ (refer to it as  the
reciprocal one).\footnote{The ultimate reason is the appearance in this gauge
of an explicit  exact solution valid also in GR with a scalar field (see
Sec.~\ref{sec:5}).} Note that as
it stands this gauge is not invariant under the
dynamical global symmetry~(\ref{gt}). Designate $A\equiv a
=1/b$, $ C\equiv r^2 c $ (some element of length $l_0$ is put to unity  here). 
 
With account~for 
\begin{eqnarray}\label{Rmn}
R_0^0&=& \frac{1}{2}\frac{(CA')'}{C},\nn\\ 
R_r^r&=&  \frac{1}{2} \frac{(CA')'}{C}
+A\bigg(\frac{C''}{C}-\frac{1}{2}\frac{C'^2}{C^2}\bigg),\nn\\ 
R_\theta^\theta&=&  \frac{1}{2}\frac{(AC')'}{C}-\frac{1}{C}
\end{eqnarray}
and $\nabla\cdot\nabla {X}=-(AC{X}')'/C$,
the  gravity equations look~like
\begin{eqnarray}\label{M0}
\label{ACs1}
CA'' + C'A'  &=& 
\upsilon_s^2(AC{X}')' -  \nn\\    - \frac{2C}{\ka_g^2}
\Big(  V_s   +  \frac{\partial V_s }{\partial  {X}} &- & 
\frac{\partial L_m}{\partial  {X}}\Big) 
 +  \frac{C}{\ka_g^2}
\bigg(T_m{}^0_0-  T_m{}^{ n}_ { n} \bigg) , \nn\\ 
\label{ACs2}
(\ln C)'' +\frac{1}{2}(\ln C)'^2
&=&- \frac{\upsilon_s^2}{2} {X}'^2 -  \frac{1}{\ka_g^2}\frac{1}{A}
\Big(T_m{}^0_0-  T_m{}^{r}_ {r} \Big) ,\nn\\ 
\label{ACs3}
CA'' - AC''  &=&-2+ \frac{2C}{\ka_g^2}
\Big(T_m{}^0_0-  T_m{}^{\theta}_ {\theta}\Big)  ,   
\end{eqnarray}
where $\upsilon_s^2\equiv2\ka_s^2/\ka_g^2$. With  $\ka_g$ taken  as
an overall   mass scale,
$\upsilon_s$ is the  single free parameter of the vacuum gravity Lagrangian 
(under $V_s=0$).  According to (\ref{M0}) 
$\upsilon_s^2 $ plays the role of a
coupling between the scalar and tensor gravities.  (For its observational
meaning  see Sec.~\ref{sec:9}.)  With $\upsilon_s\sim
10^{-3}$  the coupling proves to be   weak.  

\paragraph{Lagrange multiplier ans\"atze}

In the equations  above, the Lagrange multiplier  is excluded. It may be
revealed  through (\ref{Wla})  as follows:
\begin{equation} \label{B}
-\ka_s^2(AC{X}')'/C + \partial V_s/\partial {X}  -  \partial L_m /\partial {X}
=\la e^{-{X}} .
\end{equation}
Given a solution to (\ref{ACs1}), Eq.~(\ref{B}) defines $\la$. V.v., making
an ansatz for $\la$ (to be confirmed) one can look for a
respective solution to (\ref{ACs1}) (if any).  This allows one to (partly)
disentangle  the scalar and tensor gravity equations.

 Being interested in the case of the empty, but possibly for the
origin, space we envisage at $L_m=0$ the  three following  ans\"{a}tze.

(\/{\em i}\/)  {\em Singular ansatz}\hspace{1ex} Here $\la=\Delta_{0}$, with 
$\Delta_{0}$ being a
$\de$-type function concentrated at the origin and  determined
implicitly  through  self-consistency.

\label{S2}
(\/{\em ii}\/)  {\em Regular ansatz}\hspace{1ex} Here
$\la=\La_{0}$ everywhere (including the origin),
with $\La_{0}$ being an arbitrary constant.

(\/{\em iii}\/) {\em Interpolating  ansatz}\hspace{1ex} Here
\be\label{interpol}
\la \simeq\cases{\Delta_{0},  \ \   r <r_0,      \cr
\La_{0}  , \ \ \,  r > r_0,  }
\ee
with $r_0$ being some matching distance. In reality $\la$ should be smoothed
around $r_0$.
Consider these ans\"{a}tze  in turn (neglecting by  $V_s$).

\subsection{Quasi-harmonic solution}
\label{sec:5}

\paragraph{Reciprocal gauge}

The singular ansatz $\la=\Delta_{0}$ reduces  to the  quasi-harmonicity
condition $\nabla\cdot\nabla {X}=0$, or $(AC{X}')'=0$ (except for the origin).
The exact solution to the proper equations was
found previously in~\cite{Pir2}. Here we reproduce it in  the  nutshell.
Combining the first and the last gravity equations 
(\ref{ACs1}) one gets $(AC)''=2$ with the ensuing relation 
\begin{equation}\label{D}
AC\equiv \De=(r-r_1)(r- r_2),
\end{equation}
where the  roots $r_1$ and $r_2$ 
are a priori either real or complex-conjugate of each other. Introduce  a new
radial coordinate 
\be \label{chiD}
\chi = - \int \frac{d r}{\De } ,
\ee
with  $\chi = 1/r +{\cal O}(1/r^2)$ at $r\to \infty$ (a unit of length $l_0$ is
understood here  and in what follows).
In these terms the  first gravity equation~(\ref{ACs1}) and the scalar-field 
equation~(\ref{B})   reduce to the free  harmonic
form
\begin{equation}\label{harmon}
  d^2 \ln A/d \chi^2   =   d^2 {X}/d \chi^2 =  0,
\end{equation}
with  general solution  
\begin{equation}
   \ln A =\ln A_0 -{\nu}_0 \chi,\  \   {X}={X}_0 -\vsi_0   \chi, 
\end{equation}
depending  on  the integration constants  $  A_0$, ${X}_0$,   ${\nu}_0$ 
and~$\vsi_0$.
 Conventionally, impose the asymptotic free boundary
conditions 
$A=1$ and   ${X}=0$ at $\chi=0$ ($r\to\infty$), so that   $A_0=1$ and
${X}_0=0$. 
Substituting $\ln C=\ln \De- \ln A$ into the second gravity equation
(\ref{ACs1})
we finally find the integration condition
\be
{\nu}_0^2+ \upsilon_s^2  \vsi_0^2=(r_1-r_2)^2.
\ee
It follows that for solution to be real  it is necessary that roots  be
real ($(r_1-r_2)^2\geq 0$).\footnote{In the case
of  complex-conjugate roots
($(r_1-r_2)^2< 0$), the reality of metric (${\nu}_0^2\ge 0$) implies
$\vsi_0^2< 0$ meaning a ghost scalar-field  solution (${X}'^2<0$).}
In this case we get from (\ref{chiD}) at $r_1\neq r_2$ (let by default
$r_1\ge r_2$):
\be\label{fracharm}
\chi=- \frac{1}{r_1-r_2}   \ln \frac{r-r_1}{r-r_2}
\ee
(modulo a constant), or inversely
\be
r=\frac{r_1+r_2}{2}+\frac{r_1-r_2}{2}\, \mbox{\rm cth\,} \frac{ (r_1-r_2)  
\chi}{2}.
\ee
At $r_1=r_2\equiv r_0$ one has $\chi=1/(r-r_0)$ and, respectively,
$r=r_0+1/\chi$.
Altogether,  the respective line element and the scalar field look like
\bea
ds^2&=&q^{{\nu}}d t^2 -q^{-{\nu}}(dr^2+\De d\Omega^2),\nn\\
X&=&\vsi \ln q,
\eea
where
\be
q=(r-r_1)/(r-r_2),
\ee
with ${\nu}={\nu}_0/(r_1-r_2)$, $\vsi=\vsi_0/(r_1-r_2)$ and
${\nu}^2+\upsilon_s^2 \vsi^2=1$.

The  gravity and scalar-field equations in the reciprocal gauge, 
(\ref{M0}) and (\ref{B}), still possess  in the vacuum
a residual invariance under the global shifts of $r$. Due to the invariance 
one can redefine $r$ as $r\to r+r_2$,
so that $q=1-r_f/r$ and $\De=r^2 q$, with $r_f\equiv r_1-r_2\ge 0$.
The looked-for solution (designate it with a subscript $f$) acquires the
following standard form:
\begin{eqnarray}\label{acs}
 a_f=1/b_f&=&(1-r_f/r)^{{\nu}_f  },\nn\\
c_f&=&(1- r_f/r)^{1-{\nu}_f}, \nn\\ 
 \sqrt{2}\sigma_f/\ka_g\equiv\Sigma_f =\upsilon_s
{X}_f&=&\eta_f\sqrt{1-{\nu}_f^2}\ln
(1-r_f/r),\ \ \
\end{eqnarray}
with   ${\nu}_f= {\nu}_0/r_f$ and   a signature factor $\eta_f=\pm 1$.
The solution depends on  two canonical parameters $r_f$ and
${\nu}_f$.\footnote{A
more symmetric form of the solution (still in the gauge $ab=1$) would correspond
to the shift
$r\to r+(r_1+r_2)/2$, so that  $\chi=-(1/r_f) \ln (r-r_f/2)/(r+r_f/2)$,
$r=(r_f/2)\mbox{\rm cth\,} (r_f \chi/2)$ and $\De=r^2-(r_f/2)^2$.} 
 
For consistency  there should fulfil $|{\nu}_f|\leq 1$.
The solution  is unique (up to constants $A_0$ and ${X}_0$) and bound to be
singular, with the regular  solution being necessarily
trivial ($r_f=0$). Note that the particular case $\eta_f=1$,
${\nu}_f=\upsilon_s/\sqrt{1+\upsilon_s^2}$
 results in the relation $\ln a_f=\upsilon_s^2 {X}_f$, which proves
to correspond  to a matterless  static space structure (see Sec.~\ref{sec:8}).
The two-parameter solution above supersedes the
one-parameter Schwarzschild  solution for BHs (${\nu}_f =1$,  ${X}_f=0$).  
In GR  it describes BHs dressed with a (free massless) scalar field.\footnote{In
GR the respective solution  was first obtained  in
an implicit form in the gauge $c=1$ by  Bergmann and Leipnik~\cite{Bergmann}
(see also~\cite{Tred}). In the explicit
form (\ref{acs})  it was discovered somewhat later  in a different context  by
Buchdahl~\cite{Buch} and   was then extensively studied   in  literature  
including  modification by the  scalar-field self-interaction
(see, e.g., \cite{Jan}--\cite{Bhad3}).} 
Reflecting a singularity in space, with the coherent scalar field treated as
DM, such a scalar-modified BH may conveniently  be referred to as   the 
dark {\em  fracture}.  
Accounting for~(\ref{sigmar}) one finds    the respective 
dark modulus  as follows:
\begin{equation}\label{gurecipr}
\sqrt{- g_{\ast  f}}=(1-r_f/r)^{1-\mu_f }, 
\end{equation}
where 
\be
 \mu_f={\nu}_f+\eta_f\sqrt{1-{\nu}_f^2} /\upsilon_s .
\ee
Now, it is possible to reverse the line of reasoning by saying that the dark 
fracture   is the static space structure which originates from
the dark modulus (\ref{gurecipr}) and is given  by (\ref{acs}), with
${\nu}_f  = {\nu}_f(\mu_f)$.
In particular, to the constant modulus ($\mu_f=1$) there corresponds 
BH (${\nu}_f=1$). At $\upsilon_s\ll 1$ there fulfills
${\nu}_f\simeq1-\upsilon_s^2(\mu_f-1)^2/2$, with ${\nu}_f$ being close to
unity in a wide range of $\mu_f$. Thus, the appearance of quasi-BHs ($|{\nu}_f-
1| \ll 1$) is parametrically enhanced.
For an arbitrary fracture  the modulus, being  complex singular, can
not be brought  to $g_{\ast f}=-1$   by any real coordinate transformations.

In general, the  solution is complex interior to  $r_1$ and, as such, should be
treated here as
an analytical continuation  from the exterior region
$r>r_1$ into the complex $r$-plane with the cut $(-\infty,r_1)$. Moreover, the
event horizon at $r=r_1$ proves to be  point-like
singular~\cite{Jan}.
To avoid the complexity one could perform the admitted shift  $r\to r+r_1$, 
so that $q=1/(1+r_f/r)$ and $\De=r^2/q$.   At  $r>0$ the solution becomes  now 
real and regular, with a singularity only at $r=0$.  The same concerns the 
modulus,
which may now be brought to unity (but for $r=0$). The asymptotic of the
solution being unchanged under the shift, the latter may be considered
as a kind of ``realization''. Being mathematically equivalent,
such a ``truncated''  real form seems physically more reasonable.
Nevertheless, ultimate treating the complex singularity would be of considerable
interest.\footnote{For treating the interior of BH  as
unattainable due to its event
horizon  shrinking   to a point singularity by  dressing with  a 
massless scalar field minimally coupled to  gravity cf.~\cite{Jan}.}

Until stated otherwise, we pursue the standard form of the solution.
Decomposing the latter  in $1/r$  one gets in the leading
approximation:
\begin{eqnarray}\label{1/r'}
 a_f=1/b_f&=&1-r_g/r,\nn\\
c_f&=&1-r_c/r , \nn\\ 
\Sigma_f=\upsilon_s {X}_f&=&-r_s/r,
\end{eqnarray}
where $r_c=\sqrt{r_g^2+r_s^2}-r_g$. Here $r_g$ and $r_s$  are  two independent
phenomenological parameters 
\begin{equation}\label{chigc}
r_g= {\nu}_f r_f,\ \    
 r_s=\eta_f \sqrt{1-{\nu}_f  ^2} r_f,
\end{equation}
or inversely
\begin{equation}\label{rfbetaf}
r_f= \sqrt{r_g^2+ r_s^2}, \ \  
{\nu}_f   =r_g\Big /\sqrt{r_g^2+ r_s^2}.
\end{equation}
Taking as independent $r_f$ and $r_g$ one has 
\be
{\nu}_f=r_g/r_f,\ \  r_c=r_f-r_g,\ \ r_s=\eta_f\sqrt{r_f^2-r_g^2}.
\ee
For definiteness, we consider  the case $r_f\ge r_g\ge 0$,
with $r_s$ remaining sign-indefinite ($\eta_f=\pm 1$).
The parameters $r_g$ and  $r_s$ fix the Newtonian-Coulombic
approximation and have the meaning, respectively,
of the gravitational and scalar  radii of a fracture. 
Under the radial
rescaling $r\to \hat r(r)$, with $\hat r=r(1 + {\cal O}(1/ r))$ asymptotically, 
these parameters are invariant and may serve as a substitute to the canonical
ones, $r_f$ and ${\nu}_f$, with   $r_f$ being the  radius of
a fracture and  ${\nu}_f$  its ``nudity''.  (BHs, $\nu_f=1$,  are  the ``nude''
fractures.)

Now we may reverse  the approach. Namely, in an exterior
region where $L_m=0$  let us   look  for an
asymptotic free solution to (\ref{M0}), with $V_s=0$ and $\nabla\cdot\nabla
{X}=-(AC{X}')'/C=0$. With $\Sigma=\upsilon_s {X}$, decompose the solution  in a
power series of $1/r$. Starting from (\ref{1/r'}) with a priori arbitrary
parameters $r_g$, $r_c$ and $r_s$, we get in the second order the
restriction  $r_c(r_c+2 r_g)=r_s^2$. Afterwards we can  step-by-step uniquely 
reconstruct  the quasi-harmonic solution (\ref{acs}) in the empty
space within the series convergence region~$r>r_f$. Because we have nowhere used
any assumptions about the distribution of matter interior to $r_f$, but for
spherical symmetry, the exterior vacuum solution is to be uniquely determined
by  the  two interior  integral characteristics corresponding to $r_g$ and
$r_s$,  independent of the details of the interior
distribution. Physically,  $r_g$ reflects the net gravitating energy  of a
fracture and $r_s$    its  net systolon energy (see Sec.~\ref{sec:8}). 
If some ``reasonable''
matter distributions  exist, but for the point-like one, this  would 
extend to UBG the Birkhoff uniqueness theorem in GR. The Buchdahl
solution may thus be to dark fractures as the Schwarzschild solution is to BHs.

In the above we have in fact  shown that  
the gauge most appropriate  to  the
problem at hand  from the point of  view of its   physical content   is the
harmonic gauge. It may be given
 by requirement $\nabla\cdot\nabla \hat {X} \sim d^2 \hat {X}/d \hat r^2 $, or
in view of~(\ref{echi'})  by $(\hat a/ \hat b)^{1/2} \hat c \hat r^2=1$, with
$\hat r$ substituting $\chi$. (For completeness though, we should have found 
$\hat b$ and  $\hat c$.) The original reciprocal gauge ($b=1/a$) may, in
turn, be more
appropriate for studying the quasi-harmonic solution in the exterior region. 
Below we shortly compare the exterior solution in several
radial coordinates/gauges peculiar from supplementary points of view.

\paragraph{Astronomic gauge}

The coordinates where $\hat c=1$ are peculiar in astronomy, with  
the surface of a concentric sphere being conventionally 
$\hat S=4\pi \hat r^2$.
In view of (\ref{coordchange_g})  such a gauge results in the 
exterior coordinate transformation 
\begin{equation}
\label{astro}
\hat r =\sqrt{c(r)}\,r=r (1-r_f/r)^{(1-{\nu}_f)/2}, \ \ r>r_f .
\end{equation}
The inverse to this relation 
being given  only  implicitly, the   exact solution in these
coordinates~\cite{Bergmann} can
not, unfortunately,  be presented  in an explicit form (but for  ${\nu}_f= 1$).

\paragraph{Unimodular gauge}

The gauge $\sqrt{-\hat g} =(\hat a\hat b)^{1/2}\hat c=1$ is  peculiar by
the fact  that here the
modulus is directly reflected  by   scalar  field,
$\sqrt{-\hat g_\ast}=e^{-\hat X}$. In view of~(\ref{g(r)}) 
this gauge results in the exterior coordinate transformation
\begin{eqnarray}
\label{unimod}
&&\hat r=\Big(3\int c(r)r^2 d r\Big)^{1/3}=\nn \\
& =&\bigg(3\int (1-r_f/r)^{1-{\nu}_f} r^2 d r\bigg)^{1/3}
, \ \ r>r_f , 
\end{eqnarray}
with the integration constant  set to  zero to ensure $\hat r =r\Big(1+{\cal
O}(r_f/r)\Big)$ at $r> r_f$.

\paragraph{Transverse gauge}

This gauge is peculiar from the theoretical considerations.
It is given by   $\hat g_\ast =- 1$, with the group of
diffeomorphisms being the transverse one ($\partial_\mu
\hat \xi^\mu=0$)  in the respective coordinates. 
Otherwise, this  is equivalent to  the
gauge $\sqrt{-\hat g}=e^{\hat {X}}$ (which may also be imposed in GR with a
scalar field). Applying~(\ref{g(r)}) to $g_\ast$ one gets the  equation
\be
\frac{\hat r^2}{ r^2} \frac{d\hat  r}{d r} = \sqrt{- g_\ast
(r)} 
=\sqrt{-g(r)}e^{-{X}(r)},
\ee
so that in the exterior region 
\begin{eqnarray} \label{trans}
\hat r&=&\Big(3\int e^{-{X}(r)}c(r)r^2 d r\Big)^{1/3}=\nn\\
&=&\bigg(3\int (1-r_f/r)^{1-\mu_f} r^2 d r
\bigg)^{1/3}, \ \ r>r_f.
\end{eqnarray}
with the asymptotic relation $\hat r=r\Big(1+{\cal O}(r_f/r)\Big)$.
In gene\-ral, the exact solution can not be presented in this case in the
explicit form (including GR with a scalar field). For BHs  (${\nu}_f=\mu_f=1$)
all   the  aforementioned gauges/coordi\-nates 
(but for harmonic) coincide identically implying in particular
$g_\ast=-1$.

\paragraph{Isotropic gauge}

This gauge is peculiar from the phenome\-nological considerations.
In view of (\ref{metric}),  under the gauge $\hat c =\hat b$  one has in
the quasi-Galilean coordinates $\hat
g_{mn}=-\hat  b\de_{mn}$ ($m,n=1,2,3$), with $d s^2= a d t^2- \hat b
d\hat {\bf x}^2  $. Such a conformal  flat spatial metric
is natural  in confronting  with Newton's dynamics.
With account for (\ref{coordchange_g})  this results in the exterior  coordinate
transformation
\begin{equation}
\ln \hat r=\int \bigg(\frac{b(r)}{c(r)}\bigg)^{1/2}\frac{d r}{r}= 
\int\frac{1}{\sqrt{1-r_f/r}}\frac{d r}{r}  ,
\end{equation}
so that 
\begin{equation}\label{isotropicsol}
\hat r = \frac{r}{4}\bigg(1 +\sqrt{1- r_f/ r}\,\bigg)^2,
\ \ r> r_f,
\end{equation}
or inversely
\begin{equation}
r=\hat r\Big(1+\frac{r_f}{4 \hat r}\Big)^2,\ \ \hat r> r_f/4.
\end{equation}
This gives
\begin{eqnarray}\label{iso}
\hat a_f&=&\Big(1-\frac{r_f}{4\hat r}\Big)^{2{\nu}_f  }\Big/
\Big(1+\frac{r_f}{4\hat r}\Big)^{2{\nu}_f  } ,\nn\\ 
\hat b_f =\hat c_f&=&\Big(1-\frac{r_f}{4\hat r}\Big)^{2(1-{\nu}_f  )}
\Big(1+\frac{r_f}{4\hat r}\Big)^{2(1+{\nu}_f  )},\nn\\ 
\hat  \Sigma_f =\upsilon_s\hat {X}_f&=& 2\eta_f\sqrt{1-{\nu}_f^2} \ln
\Big(\Big(1-\frac{r_f}{4
\hat r}\Big)\Big/ \Big(1+\frac{r_f}{4 \hat r}\Big)\Big), \ \ \ \ \ \
\end{eqnarray} 
with $\hat \Sigma_f$ being the odd  function of $\hat r/r_f$.
Now the solution preserves  the dynamical global symmetry, as well as
possesses additionally a hidden symmetry: $r_f \to -r_f$,  ${\nu}_f  \to
-{\nu}_f  $
and  $\eta_f  \to -\eta_f$. The same concerns $\hat g_\ast$.

\paragraph{Post-Newtonian approximation}

Without loss of generality  present the fracture solution  in the isotropic
polar coordinates  (with
the hat-sign omitted) as follows:
\begin{eqnarray}\label{post}
 a_f&=&1-\frac{r_g}{ r}+ \frac{\beta_{f}}{2}\frac{r_g^2}{{
r}^2}
,\nn\\ 
 b_f= c_f&=& 1+\gamma_{f}  \frac{r_g}{ r}
,\nn\\ 
 \Sigma_f =\upsilon_s  {X}_f&=& - {\xi}_f \frac{r_s}{ r}.
\end{eqnarray}
Here   $\beta_f$ and $\gamma_f$ are  the effective parameters 
for a dark
fracture in  the parametrized post-Newtonian (PPN) formalism. It may be
said  in general that$(\beta-1)$ reflects empirically 
the degree of non-linearity in the superposition law for gravity, while 
$(\gamma-1)$ corresponds to the  amount of space curvature produced by  the unit
rest mass~\cite{Will}. The parameter~$\xi_f$ may be treated as an effective 
form-factor in the scalar-field Coulomb law.
Decomposing  (\ref{iso})  in  $1/  r$ and using
(\ref{rfbetaf}) one gets in the leading approximation
\begin{eqnarray}\label{post}
\beta_{f}&=&1- \Big( 1+\frac{\epsilon_f^2}{9}\Big)  \frac{3}{8}   \frac{r_g}{
r} ,\nn\\ 
\gamma_{f}&=& 1+ \Big(1-
\frac{\epsilon_f^2}{3}\Big)  \frac{3}{8}  \frac{r_g}{{ r}},\nn\\
{\xi}_f&=& 1+ 
\frac{1+\epsilon_f^2}{48} \frac{ r_g^2}{ r^2},
\end{eqnarray}
with
\be
\epsilon_f\equiv r_s/r_g=\eta_f \sqrt{1/{\nu}_f^2-1}.
\ee

At small  $\epsilon_f$  a dark fracture  closely 
reproduces BH ($r_s=0$, ${\nu}_f=1$,  $\epsilon_f=0$), with the  scalar
dressing being thus very neatly hidden.
The precision local tests of GR  in Solar System  result typically in 
$|\beta-1|\leq 10^{-4}$ and $|\gamma-1|\leq 10^{-5}$~\cite{Will}.  With  a
value near the Earth  $r_S/ r \simeq 2 \times 10^{-8}$,  $r_S$ being the
Sun
gravitational radius,  this implies for the Sun as a
dark fracture just 
$|\epsilon_S|\leq 10^2 $, or $\nu_S\ge 10^{-2}$, the
Sun scalar form-factor
${\xi}_S$ remaining  practically  unity.  At face value this gives  very loose
restriction.\footnote{The validity of decomposition
in $1/r$ implies that at the distances at hand there should take place 
$|\epsilon_f| \ll ( r/r_g)^{1/2}\simeq 10^4$, being still safely fulfilled.} 
Other  observational  manifestations are
conceivably needed to test the theory (see Sec.~\ref{sec:9}).\footnote{For
discussion of the transverse gravity vs.\  observations
cf.~also~\cite{Alvarez2}.}

\subsection{Regular non-harmonic  solution}
\label{sec:6}

\paragraph{Vacuum scaled distance}

To begin with, take $\ln A\equiv \al $ and $\ln c\equiv\zeta $ as the new
metric variables and   present the  gravity and the scalar-field
equations, respectively,  (\ref{ACs1}) and (\ref{B})
 in the empty, but possibly for the origin, space
equivalently as follows:
\begin{eqnarray}\label{lnC1}
\frac{d}{dr}\bigg(e^{\al+\zeta}\/ r^2 \frac{d}{d r} \Big (\al -\upsilon_s^2
{X}\Big)\bigg)&=& -\frac{2}{\ka_g^2}\bigg(V_s+\frac{\partial V_s}{\partial
{X}}\bigg),\nn\\
 \frac{d}{d r}\bigg( r^2\frac{d \zeta}{d r}\bigg)     
+\frac{1}{2}\bigg( r \frac{d \zeta }
{d r}\bigg)^2&=&-\frac{\upsilon_s^2}{2}
\bigg( r\frac{d{X}} {d r}\bigg)^2,\nn\\
\frac{d}{d r}\bigg( e^{\al+\zeta }\/  r^2\frac{d(\al- \zeta)}{d r}& +&
 2 r\Big(1- e^{\al+\zeta }\Big)\bigg)=0,\nn\\
e^{-\zeta}\frac{d}{d r}\bigg(e^{\al+\zeta}\/ r^2\frac{d {X}} {d r} \bigg)
&=&\frac{r^2}{\kappa_s^{2}}\bigg(\frac{\partial V_s}{\partial {X}}
-\la e^{-{X}}\bigg).\  \
\end{eqnarray}
It becomes now evident  that  the second-order system above always
has one first
integrals, with one more appearing at  $V_s=0$.  Moreover, in the latter case
the first equation above  may always be  satisfied 
with $\al=\upsilon_s^2 {X}$ (modulo a constant), with the set of solutions to
the remaining equations being definitely  not empty (cf., e.g., 
Sec.~\ref{sec:5}). 
Such a  solution proves to correspond to a matterless static space
structure (see Sec.~\ref{sec:8}).

Now, under the regular ansatz $\la=\La_{0}< 0$ everywhere
(including the origin) let us introduce in the empty space  a characteristic 
length
scale $R_{0}$ through 
\be\label{R0}
-\La_{0} =  6\ka_s^2 /R_{0}^2 =3 (\upsilon_s \ka_g  /R_{0})^2
\ee
and choose the  scaled distance $\tau=r/R_{0} $ as an independent variable.
In these terms the vacuum  scalar-field  equation  (\ref{lnC1})   at $V_s=0$  
becomes 
\begin{equation} \label{B1}
e^{-\zeta}\frac{d}{d\tau}\bigg(e^{\al+\zeta}\/  \tau^2 \frac{d {X} }{d\tau}
\bigg)=6\tau^2
e^{-{X}},
\end{equation} 
while the  tensor gravity ones  remain
unchanged modulo  substitution $r\to \tau$.

\paragraph{Approximate equations}

Of particular phenomenological interest  is the  case  $\upsilon_s \ll 1$ (see
Sec.~\ref{sec:9}). Assuming in this case ${X}=
{\cal O}(1) $ and $|\al|, |\zeta|\ll 1$ (to be confirmed) 
present the   scalar-field  and tensor gravity equations 
 in the linear in $\al$ and $\zeta$  approximation, respectively,  as follows:
\begin{eqnarray} \label{approxim}
\frac{ d}{d \tau} \bigg( \tau^2\frac{d {X}}{d
\tau}\bigg)&= &6\tau^2 e^{-{X}},\nn\\
\frac{ d}{d \tau} \bigg(\tau^2 \frac{d}{d \tau} \Big( \al-  \upsilon_s  ^2
{X}\Big)   
\bigg)
&=& 0 ,\nn\\
\frac{ d}{d \tau} \bigg(\tau^2 \frac{d \zeta}{d \tau}\bigg)
&=& -   \frac{\upsilon_s  ^2  }{2}\bigg(\tau \frac{ d
{X}}{d\tau}\bigg)^2,
\end{eqnarray}
with the last  tensor gravity  equation
\begin{equation}\label{leadcc}
\frac{d}{d \tau}\bigg(  \tau^2\frac{d }{d \tau}  (\al-\zeta)  
 - 2 \tau(\al +\zeta)\bigg)= 0
\end{equation}
serving as a  consistency condition. Clearly, the coupling of the tensor  and
scalar gravity  modes  is weak, whereas the  self-coupling of the
scalar mode  proves to be strong. The second  equation 
(\ref{approxim}) has the general solution $\al -\upsilon_s^2 {X} =
c_1/\tau +c_0$, with $c_1$ and $c_0$ being some  constants. 

The driving equation in the system above is that  for~${X}$. Having
solved the equation, one can easily find $\al$ and $\zeta$.
It follows from (\ref{approxim}) that if ${X} (\tau)$ is a particular solution 
then an equivalence  class of solutions may be obtained by the inhomogeneous
scaling transformations:
\begin{equation}\label{sim}
 {X}(\tau)\to \ti {X}(\tau)= {X}(k_0\tau )-2\ln k_0,
\end{equation}
with $k_0>0$  being  a constant.  This is a reminiscence   of 
the dynamical global symmetry~(\ref{gt{X}})  in the global
non-symmetric gauge $ab=1$. At face value,  
(\ref{sim}) reduces to reparametrization $R_{0}\to 
R_{0}/k_0$ supplemented by a shift in ${X}$.
Because $R_{0}$ is arbitrary and  a  shift in ${X}$ does not matter  at
$V_s=0$, 
we restrict ourselves by $k_0 =1$.

\begin{figure}
\begin{center}
\resizebox{0.6 \textwidth}{!}{%
\includegraphics{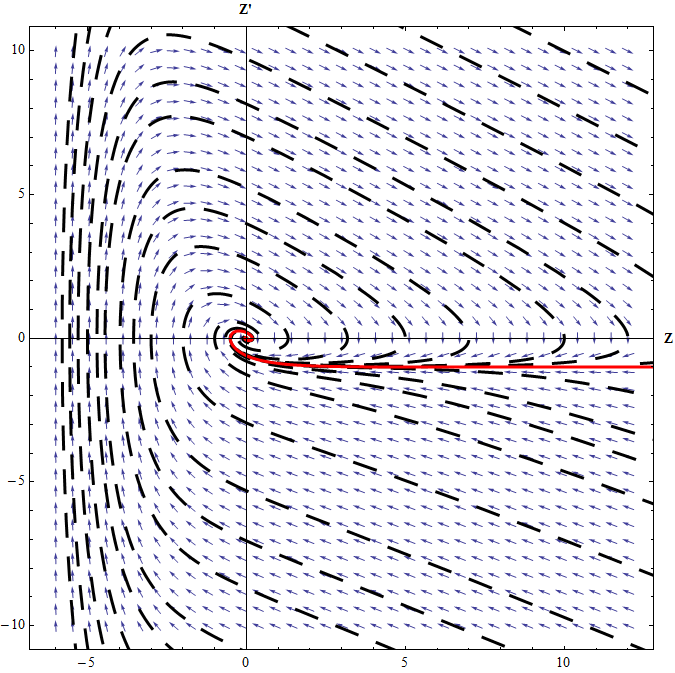}}
\end{center}
\caption{
\label{fig:1}
Phase plane $({Z},\dot {Z})$:  arrows designate the normalized
direction field determined by $d \dot {Z}/d {Z}$. The trajectories $({Z}(t),\dot
{Z}(t))$ are tangential to the direction field everywhere. Solid line -- the
regular trajectory ($\dot {Z}\to -1 $  at $t\to\ -\infty$); dashed lines --
irregular trajectories ($\dot {Z}\to -\infty $  at $t\to -\infty$). The origin
${Z}=\dot {Z}=0$ is attractor at $t\to\infty$.}
\end{figure}

To study the scalar-field  equation,  introduce the new variables
\be
t=\ln 3\tau^2,\  \ {Z}={X}-t
\ee
and present the equation equivalently as follows:
\be
\frac{d^2 {Z}}{d t^2}+\frac{1}{2} \frac{d {Z}}{d t}=\frac{1}{2}\Big(
e^{-{Z}}-1\Big).
\ee
Putting further $\dot {Z}\equiv d {Z}/d t$ one can bring the second-order
ordinary
differential equation above to the
autonomous first-order differential system:
\bea
\frac{d {Z}}{d t}&=& \dot {Z}\nn\\
\frac{d \dot {Z}}{d t}&=&- \frac{1}{2} \dot {Z}+   \frac{1}{2}\Big(
e^{-{Z}}-1\Big).
\eea
Such systems are known to be basically  characterized by the types of their
exceptional points  given by  the requirement
$d {Z}/d t=d\dot {Z}/d t=0 $. The system above has   the
single such  point, ${Z}=\dot {Z}=0$,  and the latter  proves to belong to the
stable focus type. The  respective phase plane $({Z},\dot {Z})$ is presented in
Figure~\ref{fig:1}.  It clearly shows the attraction point and a distinguished 
trajectory  (the solid line) to which all other trajectories (the dashed lines) 
tend to accumulate departing it nevertheless sooner or later.

Accordingly, there are three classes of solutions to~(\ref{approxim}):

(\/{\em i}\/)  an {\em exceptional} solution  reflected by  the  attraction
point at
the origin of  phase plane;

(\/{\em ii}\/) a {\em regular} at the origin ($\tau=0$)
solution corresponding to the regular  trajectory ($\dot {Z}\to -1 $ at
$t\to -\infty$);

(\/{\em iii}\/) the {\em irregular} at $\tau=0$ solutions corresponding to the
irregular trajectories ($\dot {Z}\to -\infty$  at   $t\to -\infty$). Consider
them in turn.

\paragraph{Exceptional   solution}

The  exact exceptional solution  corresponding to ${Z}=\dot {Z}=0$ looks like
\begin{eqnarray}
\label{except1}
\bar {X}  = \bar\al /\upsilon_s^2    &=&\ln
3\tau^2,\nn\\ 
\label{except2}
\bar\zeta/\upsilon_s^2  &=&-\ln 3\tau^2  + 2,
\end{eqnarray}
with the integration constants properly chosen. This solution may serve as a
reference one, with all other  solutions approaching  it at $\tau \to \infty$.
Note that $\bar {X}$ is invariant under the rescaling   (\ref{sim}). 
Inverting~(\ref{sigmar}) one gets in the leading
$\upsilon_s$-order  the hard-core
dark  modulus corresponding to~$\bar {X}$:
\begin{equation} \label{gcusp}
\sqrt{-\bar g_\ast} \simeq e^{-\bar {X}}=1/(3\tau^2).
\end{equation}
Implementing (\ref{g(r)})  to $\bar g_\ast$ gives  for the
transverse  coordinates  the cuspy relation $\hat  \tau= \tau^{1/3}$.

\paragraph{Regular  solution}

Decomposing a regular solution in  the powers of $\tau$ (only even powers prove
to enter)  we get the looked-for 
solution (endow it with the subscript $h$)  as follows:
\begin{equation}\label{Deltas0}
\!\!\!{X}_h= \al _h/\upsilon_s  ^2=
\cases{
\tau^2-\frac{3}{10}\tau^4
+\frac{4}{30}\tau^6 +{\cal O}(\tau^8), \   \tau\leq 1, \cr
\ln 3\tau^2, \hspace{21ex}\,   \tau\gg 1}
\end{equation}
and
\begin{equation} 
\zeta _h/\upsilon_s  ^2 =\cases{
-\frac{1}{10}\tau^4 +\frac{2}{35}\tau^6 +{\cal O}(\tau^8), \ 
 \tau\leq 1,\cr
-\ln 3\tau^2 +2, \hspace{10ex}\ \ \,  \tau\gg 1,}
\end{equation}
with $ {X} _h(0)=0$ being imposed. 
The restriction (\ref{leadcc})  fulfils  identically  with the same
accuracy  both at $\tau\leq 1$ and $\tau\gg 1$.  
Note that though ${X}_h$ changes under the global transformations (\ref{sim}) 
its asymptotic remains invariant.   The solution satisfies
the matterless condition, $\al_h=\upsilon_s^2 {X}_h$ (see Sec.~\ref{sec:8}). 
The respective static space
structure is nothing but  the dark {\em    halo}.\footnote{Indeed,
equating the Newtonian gravitational attraction force  $F_g= m \al_{h}'/2$,
acting in the given metric on a test particle with the
mass~$m$, to the centripetal force
$F_c=mv_h^2/r$, corresponding to  the circular rotation velocity~$v_h$, we
could already  anticipate the asymptotic constant
$v_h=\upsilon_s$,  characteristic of  the galaxy dark halos (see
Sec.~\ref{sec:9}).} 
The parameter $R_{0}$ plays the role of   the soft-core radius (see
Sec.~\ref{sec:9}). The equations above can be extended analytically to 
$\tau^2<0$ (corresponding to $R_{0}^2<0$, $\La_0>0$), though with loosing the 
halo-type solution. This explains the earlier made choice
$\La_{0}\leq 0$.   

Inverting~(\ref{sigmar}) one gets 
in the wide region of $\tau$, where $ | \zeta_h|={\cal O}( \upsilon_s ^2)\ll
1$,  the soft-core dark modulus as follows:
\begin{equation}\label{scmod}
\sqrt{-g_{\ast  h}} \simeq e^{-{X}_h}=\cases{1-\tau^2  +{\cal
O}(\tau^4), \ \  \tau\leq 1,\cr
1/(3\tau^2), \hspace{7ex}\ \ \    \tau\gg 1.}
\end{equation}
Implementing (\ref{g(r)}) to $g_{\ast  h}$  gives then  for the transverse 
coordinates the relation
\begin{equation}
\hat \tau\simeq\cases{\tau, \hspace{3ex}\ \tau\leq 1,\cr
\tau^{1/3}, \     \tau\gg 1.}
\end{equation}

\begin{figure}
\begin{center}
 \resizebox{0.485\textwidth}{!}{%
\includegraphics{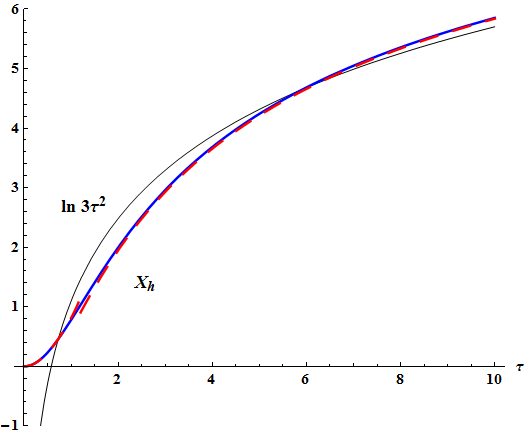}}
\end{center}
\caption{
\label{fig:2} 
Regular solution ${X}_h$: bold solid line -- numerical result; 
dashed line -- piece-wise analytical approximation  with $\bar\de_0$ and  $
\bar\tau_0$ as in Figure~3.  Thin
line --  exceptional solution.}
\end{figure}

\begin{figure}
\begin{center}
\resizebox{0.485\textwidth}{!}{%
\includegraphics{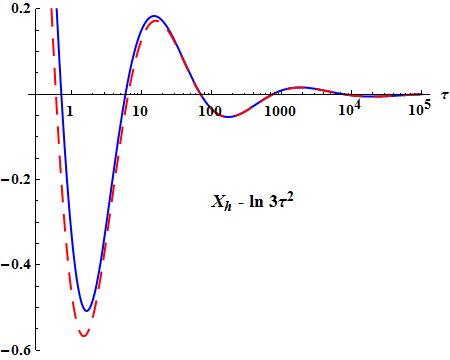}}
\end{center}
\caption{
\label{fig:3}      
The asymptotic of the regular solution ${X}_h$: solid 
line -- numerical
result for $\Delta
\bar {X}_h={X}_h-\bar {X}$; dashed line -- 
 analytical approximation with $\bar\de_0=-0.75$ and $ \bar\tau_0=2.0$ }
\end{figure}

To study the asymptotic behaviour of ${X}_h$ in more detail put    
${X} _h\equiv \bar {X}+\Delta \bar {X}_h $. In these terms  the
approximate scalar-field equation looks equivalently like
\begin{equation} 
\label{leadS''}
\frac{ d}{d \tau} \bigg( \tau^2
\frac{d\Delta \bar {X}_h}{d \tau}\bigg) = 2\Big(e^{-\Delta \bar {X}_h }-1\Big).
\end{equation}
Assuming $|\Delta \bar {X}_h |<1$ 
(to be confirmed) and retaining  the linear in $\Delta \bar {X}_h $ part
 we get the solution at  $\tau > 1$ as follows:
\begin{equation}
\Delta \bar {X}_h =
(\bar\de_0/\sqrt{\tau}) \cos\Big((\sqrt{7}/2) \ln \tau/\bar\tau_0\Big).
\end{equation}
Here $\bar\de_0$ and $ \bar\tau_0$ are some integration constants to be 
inferred from matching with solution at $\tau\leq 1$ or from comparison with the
numerical solution. Clearly, ${X}_h$ oscillates with
attenuation  around  $\bar {X}$ approaching  the latter 
at $\tau\to\infty$.
The regular solution  is unique and may be prolonged in the arbitrary order
in $\upsilon_s^2$  to the solution of the exact equations~(\ref{lnC1}).
The behaviour of ${X}_h$  is shown in Figures~\ref{fig:2}  and \ref{fig:3}.
It is seen, in particular,   that ${X}_h$ gets strong already at the
moderate~$\tau$, so that the account for potential $V_s({X})$ may become
important in this region.\footnote{For
additional details on the regular solution see~\cite{Pir3,Pir4}.}

\subsection{Irregular non-harmonic  solution }
\label{sec:7}

At last, consider  the interpolating ansatz for $\la$ given by
(\ref{interpol}). To
satisfy this ansatz,  drop off the requirement of regularity at the origin  and 
look for a  solution  interpolating between the two previous solutions. 
Put  without loss of generality ${X} = {X} _h+\Delta {X}_h $. In these terms 
the
approximate scalar-field equation corresponding    to $\la=\La_{0}$ at
$\tau>r_0/R_{0}$ looks  equivalently like
\begin{equation} 
\label{leadS'}
\frac{ d}{d \tau} \bigg( \tau^2
\frac{d\Delta {X}_h}{d \tau}\bigg) = 6\tau^2 e^{-  {X}_h}\Big(e^{-\Delta {X}_h
}-1\Big).
\end{equation}
Similarly, putting $\al= \al_h+\Delta \al_h$ and $\zeta=
\zeta_h+\Delta \zeta_h$ present the approximate gravity equations at
$\tau>r_0/R_{0}$
as follows:
\begin{eqnarray}\label{eq:Delta}
\label{leada'}
\frac{ d}{d \tau} \bigg(\tau^2 \frac{d}{d\tau} (\Delta\al_h - \upsilon_s^2 
\Delta {X}_h)\bigg)
&=&0,\nn\\ 
\label{leadc'}
\frac{ d}{d \tau} \bigg(\tau^2 \frac{d \Delta\zeta_h}{d \tau}\bigg)
&=&\nn\\ 
= -  \upsilon_s  ^2 \tau^2 \bigg( \frac{ d  {X} _h} {d\tau}
+ \frac{1}{2}  \frac{ d  \Delta {X}_h} {d\tau}  \bigg) \frac{ d  \Delta {X}_h }
{d\tau} ,
\end{eqnarray}
with the consistency condition  remaining as before:
\begin{equation}\label{leadcc'}
\frac{d}{d \tau}\bigg(  \tau^2\frac{d}{d \tau}   (\Delta\al_h-\Delta \zeta_h) 
 - 2 \tau(\Delta\al_h +\Delta\zeta_h)\bigg)= 0. 
\end{equation}

\begin{figure}
\begin{center}
\resizebox{0.485\textwidth}{!}{%
\includegraphics{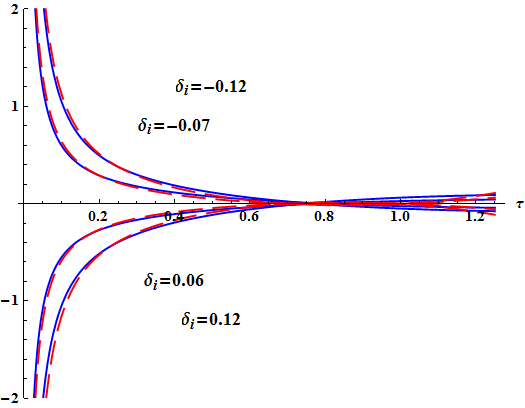}}
\end{center}
\caption{
\label{fig:4}    
Representative irregular solutions ${X}$: solid lines -- numerical
results for    $\Delta {X}_h={X}-{X}_h$; dashed lines --  analytical
approximation with the respective $\de_i$ for the  pole  singularity.}
\end{figure}

Assuming $|\Delta {X}_h |<1$ and $ |d \Delta {X}_h /d\tau|< | d {X} _h/d\tau|$
(to be confirmed) consider the linear in $\Delta {X}_h $ approximation.
Decomposing  $\Delta {X}_h $ in the series in $\tau$  starting (by
assumption to be verified) from $1/\tau$  and accounting
for (\ref{Deltas0}) we get the solution as follows:
\begin{equation}\label{lump}
\Delta {X}_h =\de_i \cases{
-1/\tau +3\tau -2 \tau^3 + {\cal
O}(\tau^5), \hspace{3ex}\,  r_0/R_{0}<\tau \leq 1,  \cr
(\de_0/\sqrt{\tau}) \cos\Big((\sqrt{7}/2) \ln \tau/\tau_0\Big), 
 \tau > 1.}
\end{equation}
In the above, $\de_i$ is a  small normalization parameter to be   fixed by
subsequent matching with the dark  fracture solution.
Likewise, $\de_0$ and $ \tau_0 $ are some
integration constants which can, in principle, be  fixed by
further applying the perturbation procedure in $\Delta {X}_h$. These constants
can be estimated by matching the two branches of (\ref{lump}) or from comparison
with the numerical results. 
Solving  (\ref{eq:Delta})  we  then  get
\begin{eqnarray}
\Delta \al_h&=&- \frac{\delta_\al }{\tau}+ \upsilon_s^2
\Delta {X}_h , \\
\Delta \zeta_h &=& -\frac{\delta_\zeta}{\tau}-
\upsilon_s^2  \de_i 
\cases{ \tau-\frac{2}{5}\tau^3+ {\cal O}(\tau^5), \
r_0/R_{0}<\tau \leq 1,\cr 
 {\cal O} (1/\sqrt{\tau}), \ \hspace{7ex} \,  \tau  >  1, }\nn
\end{eqnarray}
with $\de_\al$ and $\de_\zeta$ being some small parameters to be fixed by
further
approximation. The consistency condition  (\ref{leadcc'})  is fulfilled with the
same accuracy. Several  representative solutions corresponding to the irregular
trajectories  somewhat close to the regular one   are shown  in
Figure~\ref{fig:4}. 
The less $|\de_i|$, the better is  the approximation.\footnote{The decaying
behaviour of  the  irregular solution with $\de_i>0$  is
superseded eventually by the growing one  at the tiny $\tau$ (not shown), what
lies though beyond the region of approximation.} 

\begin{figure}
\begin{center}
\resizebox{0.485\textwidth}{!}{%
\includegraphics{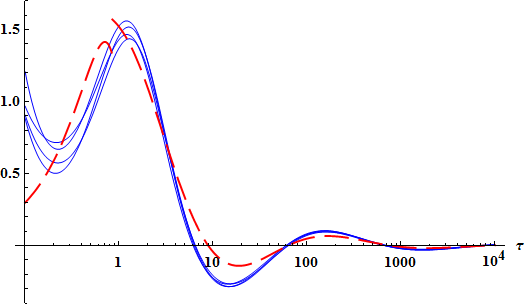}}
\end{center}
\caption{
\label{fig:5}  
Representative irregular solutions ${X}$: solid lines --  numerical 
results for  interference term $ {X}_i/\de_i$, with  $\de_i  $ as in
Figure~\ref{fig:4}; 
dashed line --  piece-wise analytical approximation,  with   $\de_0=0.85$
and~$\tau_0=2.0$.}
\end{figure}

To arrive at a solution valid approximately  in the whole interval of $\tau$,
match the $1/\tau$-pole term in 
$\Delta {X}_h$  with the respective term  in ${X}_f$, the latter corresponding  
to
$\la=\Delta_{0}$ at $\tau<r_0/R_{0}$ (and similarly for $\Delta \al_h$ and
$\Delta \al_h$). 
In view of (\ref{1/r'})  this implies
\begin{eqnarray}
 \de_i&= &r_s/ (\upsilon_s R_{0}), \nn\\
\delta_\al&=&  (r_g - \upsilon_s r_s) /R_{0}, \nn\\
\delta_\zeta&=&\Big (\sqrt{r_g^2+r_s^2}-r_g\Big)\Big/R_{0} .
\end{eqnarray}
In particular, the case $\de_\al=0$ ($r_g=\upsilon_s r_s$) 
proves to correspond to  a matterless fracture, with
$\ln a_f =\upsilon_s^2 {X} _f$ (see Sec.~\ref{sec:8}). 
Partite further  $\Delta {X}_h $ as  $\Delta {X}_h \equiv {X} _f+{X} _i$, 
where ${X} _f=-\de_i/\tau= -r_s/(\upsilon_s r)$ is the scalar tail of fracture 
and the rest,  ${X} _i$, is attributed to fracture-halo interference.
Altogether,  the total solution (designate it by the subscript~$l$) can be
presented as a coherent sum
of three contributions, ${X} _l={X}_h+\Delta {X}_h={X} _f+{X} _h +{X} _i\equiv
{X} _f+{X}_{h
eff}$ (and
similarly for the metric components $ \al_l$ and  $\zeta_l$).
At $\tau\to\infty$  the interference disappears  
due to   disappearance of   ${X}_f$.
At $ \de_i=0$ ($r_s= 0$ or $R_{0}\to \infty$)  it disappears identically.
The behaviour of  ${X}_i/\de_i$ is show in Figure~\ref{fig:5}. Finally note
that  in the Laurent decomposition of the exact 
$X_l$, the part  comprising   powers of
$1/r$ may be associated with~$X_f$, the even powers of $r$ with  $X_h$, and
 the odd powers of~$r$ with~$X_i$.

The gravitational  potential $ \al_l$ reveals
the property of confine\-ment, with  the
reciprocal behaviour,  $-r_g/r$, 
at the tail of fracture 
superseded eventually at the periphery of halo by the logarithmic
potential of  gravitational attraction,
$\upsilon_s^2 \ln 3r^2/R_{0}^2$.  The 
confine\-ment scale $R_{0}$ is otherwise the halo soft-core radius
(see Sec~\ref{sec:8}). Call
the respective static space  structure the dark {\em   
lacuna}.\footnote{The
gravitational attraction force $ m r_g/2
r^2$, acting on a test particle    with   mass $m$  inside a  lacuna  due to 
a central fracture,  is superseded eventually
by the universal  attraction force
$m \upsilon_s^2 /r$  due to halo, which should dominate at
$r>\mbox{max}(R_0,r_g/\upsilon_s^2)$.}
Due to   the  coherent scalar  field getting strong towards the
periphery of halo, the gravitational confinement within a lacuna should in
reality be only partial, being terminated ultimately   by the potential $V_s$
and/or the influence of the near-by lacunas. Generally, the lacunas depend on
three distance scales which may be chosen as $r_g$, $r_s$ and~$R_{0}$.
With  the  inner and outer scalar distribution  radii, $r_s$ and~$R_{0}$, 
getting closer, the three-component structure of a lacuna becomes less
prominent. Nevertheless,  the asymptotic behaviour and 
gravitational confine\-ment  in a lacuna should still survive (at least in a
parameter region). This is  because the
asymptotic properties of the solution are determined exclusively by the
exponential term in the r.h.s.\ of the vacuum scalar-field 
equation   (\ref{B1}) or  (\ref{approxim}).

\section{~Interpretation and applications}
\subsection{Energy content}
\label{sec:8}

\paragraph{Static space structures}

The preceding results were obtained exclusively in the geometry framework
without any recourse to DM. To the latter end  let us reveal the energy
content of the static space structures  found previously. 
Consider an isolated gravitationally bound system. Let
${T}_{\mu\nu} $ be its total bare  energy-momentum 
tensor,   with gravity  included  only in the
minimal fashion through  metric. To  incorporate the  proper gravity
contribution  define the net energy-momentum pseudo-tensor 
$\check T_{\mu\nu} ={T}_{\mu\nu} +
\check\tau_{\mu\nu}$,
with $\check\tau_{\mu\nu}$ being the pseudo-tensor of  the gravity itself. At
that,
$\check\tau_{\mu\nu}$ is well-known to be dependent 
both on  definition and, possibly,  the class of coordinates. The same  concerns
thus~$\check T_{\mu\nu}$. According to
Tolman~\cite{Tal}, with   $\check\tau_{\mu\nu}$ taken as  the Einstein
pseudo-tensor in the quasi-Galilean coordinates,   the net gravitating 
energy/mass of a static isolated system may be  expressed entirely  through
its bare  energy-momentum tensor  as follows:
\begin{eqnarray}
M_g&=& \int \check T{}^0_0\sqrt{-g}d^3 x=
\int ( {T} ^0_0-{T}_l^l )\sqrt{-g}d^3 x.
\end{eqnarray}
In other terms, the net energy density, with  gravity properly 
incorporated, is $ \check \rho \equiv \check T{}^0_0= \rho+\Sigma p_l $, 
where  $\rho$ and
$p_l$  ($l=1,2,3$) are the bare energy  density and
pressure, respectively. 
Physically this means that gravity ``pumps-in''   energy   equal 
to the proper work performed  against pressure.
Having been established in the particular
spatial  coordinates, $M_g$
is to be prolonged to the arbitrary ones as a  scalar. 
For validity of the isolation assumption,  the ever-present cosmological
constant $\Lambda$, which is paramount for the Universe as a whole,  is to be
excluded from $\check\rho$   for an isolated  system. (Otherwise this would
result in a double-counting when considering the evolution of the Universe.)
In view of (\ref{R00}) we get most generally 
\begin{equation}\label{M2}
M_g= 2\ka_g^2 \int R^0_0\sqrt{-g}d ^3 x,
\end{equation}
with the net gravitating energy  determined thus entirely by the
$R^0_0$-component of the Ricci tensor.

In the reciprocal gauge, by means of (\ref{Rmn}) and the
first equation  (\ref{echi'})
(with $1/b=a\equiv A$ and   $\si$ substituted by the spatial scalar $\ln A$)
we
get  
\begin{equation}\label{R00A}
2 R^0_0= (AC(\ln A)')'/C= -
g^{kl}\nabla_k  \nabla_l \ln a.
\end{equation}
Here $\nabla_k$ is a spatial component of the four-dimensional
covariant derivative. This relation is now valid in
the arbitrary spatial gauge/coordinates. 
This gives
\begin{equation}\label{m}
M_g= - \ka_g^2\int  \sqrt{-g} g^{kl}\nabla_k \nabla_l  \ln a\, d ^3 x.
\end{equation}
In view of (\ref{d'Alambert}) (with   ${X}$ substituted by $\ln  a$) one has
\begin{equation}
\sqrt{-g}g^{kl}\nabla_k  \nabla_l \ln a= \partial_k
(\sqrt{-g}  g^{kl} \partial_l \ln a),
\end{equation}
and the three-dimensional Gauss theorem reduces (\ref{m}) to the
integral over the remote two-dimensional surface $S$ as follows:
\begin{equation}
M_g=-\ka_g^2 \int  \sqrt{-g}g^{kl} \partial_l \ln a\, d S_k,
\end{equation}
with $\sqrt{-g}=\sqrt{ab}c$.  

Partite  $M_g$  onto the contributions of the non-gravita\-tional matter and
the systolons, respectively, $M_m$ and $M_s$, including the proper
gravity energy:
\begin{eqnarray}
\! \! M_g&=& M_m+M_s =\nn\\
&= &\int
\Big((T_m{}^0_0-T_m{}^n_n)+  ({T}_{s}
{}^0_0-{T}_{s} {}^n_n) \Big)\sqrt{-g}\,d^3 x.\  \
\end{eqnarray}
Put by default  for the net partial energy densities:
\be 
\check \rho_{m}\equiv  T_m{}^0_0-T_m{}^n_n,\ \
\check \rho_{s}\equiv   T_s{}^0_0-    T_s{}^n_n.
\ee
In view of (\ref{trace})   the systolon net partial  energy  (with 
$V_s$ neglected) is
\begin{equation}\label{mh}
M_s= -  \upsilon_s^2\ka_g^2 \int  \sqrt{-g} g^{kl}\nabla_k \nabla_l  
{X}d ^3 x, 
\end{equation}
and the  three-dimensional Gauss theorem  gives then 
\begin{equation}\label{Mh}
M_s =- \upsilon_s^2\ka_g^2\int   \sqrt{-g}g^{kl} \partial_l  {X} d S_k.
\end{equation}

Treated as general covariant scalars the expressions above  determine  in UBG
the energy
content of a static space structure  through
$\nabla\cdot \nabla \ln a$ and
$\nabla\cdot \nabla {X}$. While the first term is the same as in GR with a
(free massless) scalar field,  the second term is peculiar to  UBG. It is
produced due to the unimodularity constraint 
missing in GR.\footnote{Physically, the scalar-graviton field  may be attributed
 to (singularity of) modulus  as a  scalar source. This may take place even in
vacuum.  In GR, to reproduce $M_s$ one should  couple the  scalar field   
directly to a matter scalar charge. Thus, though the solution in the two
theories is formally the same, its physics content differs~\cite{Pir2}.} 
In particular, in the case $\ln a = \upsilon_s^2 {X}$ (modulo an
additive constant)  there follows  $M_g=M_s$,   $M_m=0$ signifying  a
matterless,  pure systolon static space structure.

\paragraph{Dark fractures}

In what follows we adopt for a dark fracture  the   truncated real form, with a
singularity only at $r=0$ (see Sec~\ref{sec:5}). Everywhere, but for the
origin, there fulfills 
$R^0_0=0$ and $\nabla\cdot\nabla  {X}=  0$.  
Thus, the spatial integrals are saturated in this
case exclusively at
$r=0$ by the point-like singularity. 
The latter being in fact known only implicitly, a spatial integral may be
substituted   via the  three-dimensional
Gauss  theorem   by the respective integral  over the remote sphere. Having
consistently defined the energy in the quasi-Galilean coordinates, one
can use,  just for calculations,  the polar coordinates, with 
$\partial_{ l}= \delta_{l}^r \partial_r$,
$\sqrt{-g}d S_k =\delta^r_{ k} c r^2 d\Omega^2$ and
$g^{rr}=-A$. The net gravitating  energy of a fracture  is then as follows:
\begin{equation}\label{M1}
 M_g=  4\pi \ka_g^2 {\nu}_f   r_f = r_g/(2G)\ge 0 ,
\end{equation}
with $r_g\ge 0$ (${\nu}_f\ge 0$) imposed.
Likewise, the  systolon net contribution is
\begin{equation}\label{m_fhi1}
 M_s=  4\pi \ka_g^2
\upsilon_s  \eta_f\sqrt{1-{\nu}_f  ^2}r_f 
 = \upsilon_s r_s/(2G),
\end{equation}
or otherwise, $M_s = \upsilon_s \eta_f\sqrt{1/{\nu}_f^2-1} M_g $. 
The net contribution of the non-gravitational matter  is then
\begin{equation}
M_m=M_g-M_s=
\Big(1-\upsilon_s\eta_f  \sqrt{1/{\nu}_f  ^2-1}\Big) M_g.
\end{equation}
Imposing by default the  requirements $M_m\ge 0$ and $M_s\ge 0$ ($\eta_f =1$) 
one  can envisage  two following extreme cases.

(\/{\em i}\/)  {\em BHs\/}: $ {\nu}_f =
{\nu}_{f{\rm max}}=1$, with $M_s=0$ and $M_g=M_m$. Here $r_g$ is
determined conventionally by the matter net energy, $r_g=2G M_m$.

(\/{\em ii}\/)   {\em Vacuum dark fractures\/}: 
${\nu}_f  = {\nu}_{f{\rm min}}  =\upsilon_s/\sqrt {1+ \upsilon_s^2} $,
with   $M_m=0$ and $M_g=M_s$.  Here $r_g= \upsilon_s r_s$, with $r_g \ll r_s$
at $\upsilon_s\ll 1$. 
This  accords with  $\ln a_f=\upsilon_s^2 {X}_f$ (modulo a constant) as a 
generic condition  of the absence of  matter.\footnote{ The gravitational
attraction of  the ``normal'' fractures ($\eta_f=1$, $r_s\ge 0$) is enhanced,
$M_g\ge M_m$.
Admitting $\eta_f=-1$, relaxing  thus  requirement
$M_s\ge 0$, one would get 
the ``anomalous''  fractures  screening the matter, $M_g< M_m$.}

\paragraph{Dark halos}

\begin{figure}
\begin{center}
\resizebox{0.485\textwidth}{!}{%
\includegraphics{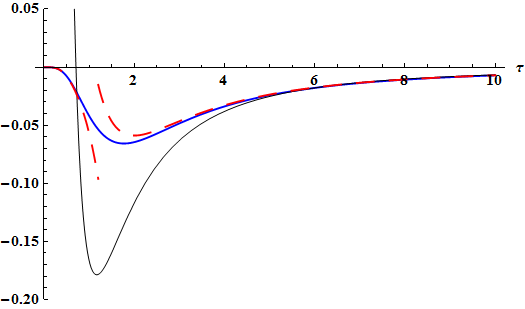}}
\end{center}
\caption{
\label{fig:6}  
Normalized energy density profiles $\rho(\tau)/\rho_{0}$: the 
difference between a  profile and the reference  one
$\rho_{ref}/\rho_{0} =1/(1+\tau^2)$. Bold solid
line -- numerical result for the soft-core profile $\rho_h/\rho_{0}$;
dashed line --  piece-wise analytical approximation;  thin line --   the
hard-core profile $\bar \rho/ \rho_{0} =1 /3\tau^2$ for exceptional solution.}
\end{figure}

For a dark halo the solution is regular, and according to (\ref{trace})  
the systolon net    energy density  is well-defined~as
\be\label{rho_s1}
 \check \rho_{s} =-2 {\La}_s=-2\La_{0} e^{-{X}}.
\ee
Under the attractive effective  potential $ \La_s$ ($\La_0<0$) this energy
density is positive-definite, with $\check \rho_s=0$ only  at $\La_0=0$.
In view of (\ref{R0}) and the first equation (\ref{approxim}) one has 
\begin{equation}\label{checkrho}
\check \rho_{ s}=\upsilon_s^2 
\frac{  \ka_g^2}{R_{0} ^2}
\frac{1}{\tau^2} \frac{d}{d\tau}\bigg(\tau^2 \frac{d {X}}{d\tau}\bigg).
\end{equation}
Substituting  ${X}= {X}_h$ one then gets in particular the halo energy density
profile as follows:
\begin{equation}\label{DM'}
\rho_{h} = \rho_{0}\cases{1-\tau^2
+\frac{4}{5}\tau^4  
+{\cal O}(\tau^6), \ \   \tau\leq 1, \cr
1/(3\tau^2)  +{\cal O}(1/\tau^{5/2}) ,  \hspace{2ex}\   \tau\gg 1,}
\end{equation}
with  $ \rho_{0}$ standing for the  central energy density:
\begin{equation}\label{DM1}
 \rho_{0}=-2\La_0= 6 \upsilon_s^2\frac{\ka_g^2}{R_{0}^2}.
\end{equation}
Thus, a dark halo naturally enjoys the soft-core energy density profile, with
$R_{0}$ being the  halo core radius. 
At that, $  \rho_h/\rho_{0}$  closely reproduces  
the dark modulus   (\ref{scmod}). 
For comparison, the exceptional solution $  \bar {X}(\tau)=\ln 3\tau^2$ 
results in  the hard-core profile
\begin{equation}
\bar \rho= \frac{1}{3} \frac{\rho_{0}}{\tau^2}
= 2\upsilon_s  ^2 \frac{\ka_g^2 }{r^2}
\end{equation}
reproducing, accordingly,   hard-core modulus (\ref{gcusp}).
Asym\-ptotically,
$\rho_{h}(\tau)$ oscillates  around
$\bar\rho$ with attenuation, approaching $\bar\rho$  at $\tau \gg 1$.  
The normalized difference, $(\rho-\rho_{ref})/\rho_0$,  
between a profile and the reference one,
$\rho_{ref}=\rho_{0}/(1+\tau^2)$, is shown   at the moderate~$\tau$     in 
Figure~\ref{fig:6}.

It is seen, in particular,  that  the soft-core profile, $\rho_h$, 
decays at  $\tau\leq 2$  faster than the reference one,
both having  the same behaviour near the
origin.\footnote{In the conventional DM approach  $\rho_{ref}$ is due to the
so-called pseudo-isothermal sphere, while $\bar \rho$ corresponds to the true
isothermal one (see, e.g.,~\cite{Blok}).}

The halo net gravitating energy interior to~$\tau$ is 
\begin{equation}
M_{h<}(\tau)=8\pi \upsilon_s^2   \ka_g^2 R_{0}
\cases{\tau^3 -\frac{3}{5}\tau^5  
+{\cal O}(\tau^7), \   \tau \leq 1  , \cr
\tau+{\cal O}(\sqrt{\tau}),  \  \hspace{7ex}
\tau\gg1.} 
\end{equation}
At $r\gg R_{0}  $,  one has $ M_{h<}(r)= 8\pi  \upsilon_s^2 \ka_g^2
  r =\upsilon_s  ^2 r/G$, with the respective energy for  
 $ \bar {X}$  rising linearly identically. 

To clarity the physics content of a dark halo  note that according  to
(\ref{rhop}), (\ref{R0}) and (\ref{rho_s1}) 
the exceptional solution  in  the leading
$\upsilon_s$-approxima\-tion results in 
 the bare  energy-momen\-tum tensor with
\begin{equation}
\bar p_0=\bar p_\theta =\bar p_\varphi =0 ,\ \
\bar p_r=  \frac{1}{3}\frac{\rho_0}{\tau^2}.
\end{equation}
In turn, this results in  the positive-definite   net energy density $\bar
\rho=\bar
p_0+\Sigma
\bar
p_n=\bar p_r=\rho_0/(3\tau^2)>0$,  being attributed entirely to
tensor gravity. As for the regular solution  ${X}_h$   (${X}'_h|_{r=0}=0$), its
bare energy density
 (\ref{rhop}) near the origin
is negative, while the  net one   (\ref{rho_s1}) is nevertheless
positive-definite everywhere ($\La_0<0$). The same concerns the bare radial
pressure~(\ref{rhop}).
At $\tau\gg 1$ the regular solution behaves as $\bar {X}$, with the
net energy density decaying like $1/\tau^2$, too.  Thus, the dark halo is
a  gravitationally tightly bound structure, with the bulk of its net
energy provided by  tensor gravity to  form the halo. 

\paragraph{Dark lacunas}

\begin{figure}
\begin{center}
\resizebox{0.485\textwidth}{!}{%
\includegraphics{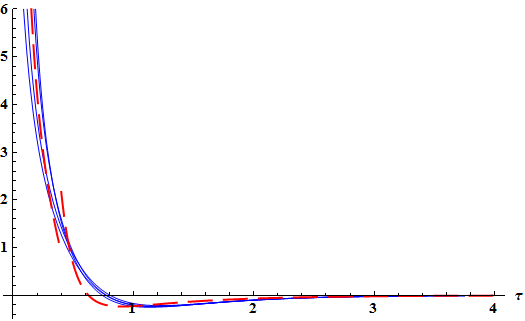}}
\end{center}
\caption{
\label{fig:7}   
Normalized energy density profiles $\rho(\tau)/\rho_{0}$: solid lines --
numerical results for interference term $\rho_i/(\rho_{0}\de_i)$,  
with $\de_i$ as in Figure~4;  
dashed line --   piece-wise analytical approximation, with $\de_0$ and
$\tau_0$ as in Figure~5.}
\end{figure}

For a dark lacuna one has
${X}_l= {X}_f+{X}_h+{X}_i\equiv {X}_f+{X}_{h eff}$. 
The  contribution  to the net gravitating  energy due to  ${X}_f$ is accounted 
for by the
surface integral.  For   energy density  of
the effective  dark halo,
$\rho_{h eff}= \rho_h +\rho_i$,   Eq.~(\ref{checkrho}) gives
 $\rho_h$ as before and the interference contribution as
follows: \begin{equation}
\rho_i/\rho_{0} =\de_i 
\cases{1/\tau-4\tau + {\cal O}(\tau^3), \ \   \tau \leq 1,\cr 
{\cal O}(1/ \tau^{5/2}), \hspace{8ex} \ \, 
\tau \gg 1.}
\end{equation}
Thus, $\rho_{h eff}$  comprises some  cuspy
$1/\tau$-correction compared to the soft-core profile
$\rho_h$.
For  the normal   fractures
($\eta_f=1)$  the interference is constructive,  $\de_i>0$, with the
effective halo  net energy 
near the origin  increasing.
The  numerical results for the  normalized interference term,
$\rho_i/(\rho_{0}\de_i)$,  are shown in Figure~\ref{fig:7}.
The interference contribution to the lacuna  net  energy 
 interior to $\tau$~is
\begin{equation}
M_{i<}(\tau)= 12\pi \upsilon_s  \ka_g^2 r_s  
\cases{\tau^2-2\tau^4 + {\cal O}(\tau^6), \   \tau \leq 1,\cr
{\cal O}(\sqrt{\tau}), \hspace{11ex}\,  \tau \gg 1.}
\end{equation}
The above  results are obtained in the perturbative fashion. For the realistic 
lacunas the picture may become   more complicated  in detail  though its
salient  features should, conceivably, survive. 

\subsection{Rotation curves and equivalent DM}
\label{sec:9}

\paragraph{Rotation  velocity}

Let $U^\mu=d x^\mu/d \tau$,  $U\cdot U=g_{\mu\nu} U^\mu
U^\nu=1$, be the
four-velocity of a test particle with respect to
its proper time $\tau$.  If the particle interacts gravitationally only in a
minimal fashion through metric, its 
 $U^\mu$ for the free motion satisfies the geodesic equation,
$\label{Ga} d  U ^\lambda/d \tau+ \Gamma^\lambda_{\mu\nu} U ^\mu
 U ^\nu=0$,
with $\Gamma^\lambda_{\mu\nu}$ standing for the Christoffel
connection. For the circular rotation in a static
spherically symmetric metric one has in the polar coordinates $ U ^r=0$.
Besides, one can  put $\theta=\pi/2$, with
$ U ^\theta=d  U ^\theta/d \tau= 0$. The remaining non-zero components  are 
$ U ^\varphi= d \varphi/d \tau$ and $  U ^0= d t/d \tau$.  The ratio
$\omega= U ^\varphi/  U ^0=d \varphi/d t$ is thus
nothing but the angular velocity  with respect to the observer's time. 
The visible   circular rotation velocity is as follows: 
\begin{equation}
u=\sqrt{cr^2}  d \varphi/d
t=\sqrt{cr^2}U^\varphi/ U^0. 
\end{equation}
For the spherically symmetric metric (\ref{polc}) one has
$\Gamma^r_{00}=a'/(2b)$,
$\Gamma^r_{\varphi\varphi}|_{\theta=\pi/2}=-(cr^2)'/(2b)$ and
$\Gamma^r_{r\varphi}\equiv0$, so that the equation of motion gives $d U^r/d
\tau=-(\Gamma^r_{00}(U^0)^2+  \Gamma^r_{\varphi\varphi}(U^{\varphi})^ 2) =0$,
and thus
\be\label{Uphi/U0}
(U^{\varphi })^2/(U^{0 })^2= a'/(cr^2)' .
\ee
Altogether, the visible rotation velocity squared is 
\be
u^2= a'/\ln (cr^2 )'.
\ee
Otherwise, the rotation  velocity with respect to the particle proper time
looks  like
\begin{equation}
V=  ud t/d \tau=u U^0=\sqrt{cr^2} U ^\varphi. 
\end{equation}
Accounting for     $U\cdot U=a (U^{0})^2- cr^2
(U^{\varphi})^2=1$  and (\ref{Uphi/U0}),
one then gets $V=v/\sqrt{1-v^2}$,  where 
\be\label{v2lna}
v^2= (\ln a)'/(\ln cr^2)'.
\ee
In the above, $v\equiv u/\sqrt{a}=V/\sqrt{1+V^2}$ is the visible rotation
velocity
accounting for the gravitational  deceleration of time.   For consistency,
$v<1$.
Being gauge invariant, all these expressions are valid in the arbitrary radial
coordinates.  The choice of the expressions  for the rotation velocity 
remains though convention dependent. By default, we choose~$v$. Particularly, 
in the astronomic coordinates
($c=1$) one gets conventionally $v^2=r(\ln a)'/2$. In the non-relativistic 
weak-field limit, we are interested in, the difference between the definitions
becomes irrelevant.\footnote{In passing, choosing $u$ one would get that  for
BHs ($a=1-r_g/r$, $ c=1$)  in all  the aforementioned radial coordinates (see
Sec.~\ref{sec:5}), but for isotropic ones, there exactly fulfills the third
Kepler's law: $T^2\sim r^3$, with $T$ being the rotation period.}

\paragraph{Rotation curves}

\begin{figure}   
\begin{center}
\resizebox{0.485\textwidth}{!}{%
\includegraphics{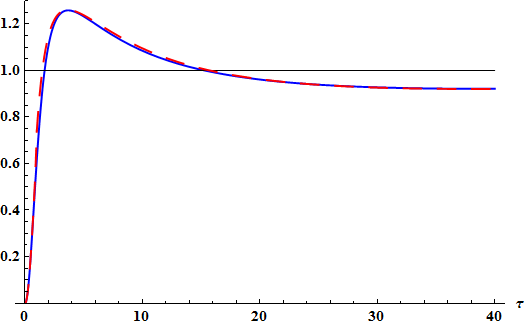}}
\end{center}
\caption{
\label{fig:8}    
Normalized RC profile $v(\tau)$: bold solid line --
numerical result for $v_h^2/\upsilon_s^2$ practically
coinciding with   analytical
approximation (dashed line);   thin
line --   ($\bar v^2/\upsilon_s^2=1$) for exceptional solution.}
\end{figure}

\begin{figure}
\begin{center}
\resizebox{0.485\textwidth}{!}{%
\includegraphics{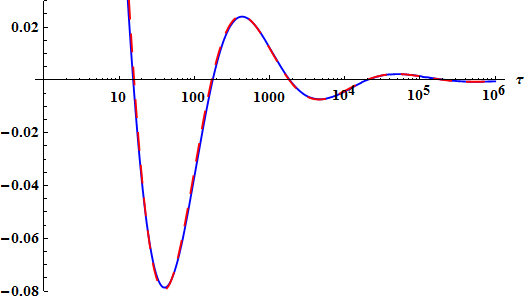}}
\end{center}
\caption{
\label{fig:9}    
The asymptotic of the normalized RC profile
$v(\tau)$: solid line --
numerical result for ($v_h^2/\upsilon_s^2-1$)  practically coinciding with  
analytical approximation (dashed line).}
\end{figure}

\begin{figure}
\begin{center}
\resizebox{0.485\textwidth}{!}{%
\includegraphics{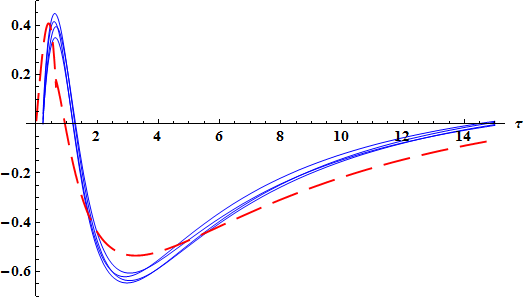}}
\end{center}
\caption{
\label{fig:10} 
Normalized RC profile $v(\tau)$: solid lines --
numerical results for interference term  $v_i^2/(\upsilon_s^2\de_i)$, with
$\de_i$ as
in  Figure~4;  
dashed line --  piece-wise analytical approximation, with $\de_0$ and $\tau_0$
as in Figure~5.}
\end{figure}

In the leading  $\upsilon_s$-approximation  one can put $c=1$, so that
\begin{equation}\label{v2}
v^2   = r\al'/2=(\tau/2) d \al/ d \tau .
\end{equation}
The total velocity squared  in a dark lacuna is thus the sum of
three components, $v^2_l=v^2_f +v^2_h+v^2_i\equiv v^2_f +v^2_{h eff}$,
where the fracture, halo and interference contributions, respectively,   are as
follows:
\begin{eqnarray}\label{vi2}
v_f^2&=&   \frac{1}{2} \frac{r_g}{r}, \    \hspace{24ex} \ \ \, r\gg r_g, \nn\\
v_h^2&=&\upsilon_s^2
\cases{\tau^2 -\frac{3}{5}\tau^4 +\frac{12}{35} \tau^6 
+ {\cal O}(\tau^8), \   \tau \leq 1,\cr
1+{\cal O}(1/\sqrt{\tau}), \hspace{12ex} \   \tau \gg 1,}\nn\\
v_i^2&=& \frac{3  }{2}   \upsilon_s^2 \de_i
\cases{\tau- 2\tau^3 +{\cal O}(\tau^5),  \hspace{5ex}\ \,  \tau
\leq 1,\cr
{\cal O}(1/\sqrt{\tau}), \hspace{12ex} \ \   \tau\gg 1, }
\end{eqnarray}
with $v_i^2$ being  generally sign-indefinite.
It is seen  that the RC profile
$v_{h eff}(\tau)$ for the effective  halo  gets   flat
only asymptotically. 
At that, the exceptional solution $\bar {X}$
would result in the precisely flat profile
$\bar v= v_{h\infty}  =\upsilon_s$,
around which  all the profiles $ v_{h eff}(\tau)$  oscillates
with attenuation, approaching $\bar v$ at $\tau\gg 1$. 
The respective results are shown in Figures~\ref{fig:8}--\ref{fig:10}.
Evidently, the net velocity profile is in reality far from being
exactly flat.

\paragraph{Equivalent non-relativistic DM}

Let us interpret the  previously found energy density and rotation velocity
profiles in terms of the non-relativistic  DM,
the latter being conventionally used in confronting   with
astronomic observations. 
For the non-relativistic DM halo  in the flat space-time ($a=c=1$) the Newton's
dynamics 
results in the rotation velocity $v_{d}$ determined through 
\begin{equation}\label{N}
v_{d}^2/r=G  M_{d< }(r)/r^2,
\end{equation}
where $ M_{d<}(r) =4\pi\int_0^r  \rho_{d}(r) r^2d r$ is
the DM energy interior to $r$, with  $ \rho_{d}$ being the
DM energy density. The latter  should thus satisfy
\begin{equation}\label{rhov}
\rho_{d}=\frac{1}{4\pi G} \frac{(r v_{d}^2)'}{r^2}.
\end{equation}
Imposing $v_{d}^2 \equiv v_{h eff}^2  = v_h^2+
v_i^2$ and accounting  for (\ref{v2})
we get the energy density  for the equivalent DM as follows:
\begin{equation}\label{rhoa}
\rho_{d}= \frac{\ka_g^2}{R_{0}^2}   \frac{1}{\tau^2}
\frac{d}{d \tau}\bigg(\tau^2\frac{d \al_{h eff} }{d \tau}  \bigg),
\end{equation}
where $\al_{h eff}= \al_{h}+\al_{i}$.
Further, because in the second line  of (\ref{approxim}) the term $1/\tau$ due
to fracture drops off, one gets hereof  $\al_{h eff}= \upsilon_s^2 {X}_{h eff}$,
and  (\ref{checkrho})  for $X_{h eff}$ gives $ \rho_{d}=\rho_{h eff} =
\rho_{h}+\rho_{i} $. 
The  profiles $\rho_{d}(\tau)$ oscillate with attenuation
around $\bar\rho $ approaching  the latter at $\tau \gg 1$ 
(cf., e.g.,~Figure~9).

Thus in the context  of
RCs, the previously found energy content of the  dark halo  coincides in
a consistent manner with the non-relativistic DM interpretation.
At that, RCs admit a complementary approach: either directly in terms of
the static space structures in  the geometry framework, without any
recourse to DM,   or in terms of the equivalent non-relativistic DM in the
framework of the Newtonian dynamics. This
justifies  treating  the coherent  systolon field as DM. 

For more clarity let us present $\rho_d$  as follows:
\be
\rho_{d} = (\rho_0/6) \chi^4 d^2{X}_{h eff}/d\chi^2 ,
\ee
where $\chi =1/\tau$, $\rho_0$ is given by (\ref{DM1}) and there roughly
fulfils ${X}_{h eff}\simeq {X}_h$, with
\be
{X}_h= \cases{
1/\chi^2, \ \ \ \ \ \ \,   \chi\ge 1,&\nn\cr
\ln (3/\chi^2), \ \ \chi\ll 1,}
\ee
so that
\be
\rho_d\simeq \rho_0\cases{
1, \ \ \ \ \ \ \,   \chi\ge 1,&\nn\cr
\chi^2/3, \ \ \chi\ll 1.}
\ee
(Note in parentheses that $\chi$ is nothing but the radial harmonic coordinate
for the dark halo, with the approximate scalar-field equation
(\ref{approxim}) being $d^2 {X}_h/d \chi^2 =(6/\chi^4) e^{-{X}_h}$. Also, under
the choice of the unit  of length $l_0=R_0$ the
fracture  harmonic  coordinate (\ref{fracharm}) smoothly  matches at $r_f<
r< R_0$  with $\chi$  for halo.)
Evidently,   it is only the relatively rapidly varying with $\chi$ part of
the scalar-graviton  field in halo   which  matters for  the equivalent DM.
At that,  varying slower  than logarithm   (though, possibly, larger)  part of 
the field may result in the
equivalent DE through the potential $V_s({X})$ (omitted here). The spherical
symmetry being not crucial, we expect that this is a generic property of the
systolon DM and DE.

We believe that  such a coherent scalar-graviton   field is of principle
importance only
within the gravitationally bound systems, such as  galaxies and clusters of
galaxies. Beyond them  the specific nature of the interior DM may become less 
important, with  the Universe
as a whole  being described effectively   by  the conventional $\La$CDM model
(or a variation of  it).
This would  result in the apparent 
two-component~DM and could, perhaps,  help unravelling the so-called
core/cusp DM  problem~\cite{Blok}.\footnote{Besides, as an  
incoherent  component of  DM  built of systolons there might serve   the (near)
vacuum mini-fractures (see Sec.~\ref{sec:8}) distributed all over the
Universe.} 

\paragraph{Galaxy DM: coherent scalar field vs.\ continuous medium}

To refine the nature of  the equivalent DM let us  consider an 
isotropic continuous medium  with the conventional 
energy-mo\-men\-tum tensor 
\be
T_d^{\mu\nu} = (\rho_d+p_d) U_d^{\mu} U_d^{\nu} - p_d g^{\mu\nu},
\ee
where $\rho_d$ and  $p_d= p_d(\rho_d )$  are, respectively,   the proper 
energy density and  pressure of the putative medium, and $U_d^{\mu}$ ($U_d\cdot
U_d=1$)  is its local four-velocity. According to~(\ref{R00}), in the
comoving frame, $U_d^{\mu}= (1,0,0,0) $,   (assumed to coincide with that of 
galaxy)   in the   metric (\ref{polc})  there fulfills: 
\be 
R_0^0=\frac{1}{2 \kappa_g^2}\Big( a\rho_d + (a+2) p_d\Big).
\ee
Imposing in the weak-field  ($|a-1|\ll 1$) approximation
\be\label{cfcm}
\rho_d+3p_d=  \rho_{h eff},
\ee
 i.e., equating the net energy densities    one reproduces the same  $R^0_0$ for
 the medium and the effective halo.
With $R_0^0$ largely determining $\ln a$ through (\ref{R00A}), and $\ln a$, 
in turn,  largely describing RCs thro\-ugh (\ref{v2lna}), this ensures   the
equivalence between the two
descriptions, the coherent-field and continuous-medium on\-es, in the
non-relativistic RC context.  Under (\ref{cfcm}) such an  equivalence fulfills
irrespective of the equation of state  $p_d =p_d( \rho _d) $.  
 In particular, at
$p_d\ll \rho_d$ one recovers the condition of the preceding paragraph,
$\rho_d=\rho_{h eff}$, for the cold/warm  DM. 
As far as being determined largely by
the Newtonian potential,  $\ln a$, the same statement  concerns  other 
gravitational manifestations  in the non-relativistic weak-field
approximation.

\paragraph{Soft-core DM halos}

The observational data on the DM-dominated (late-type LSB disk  and
gas-rich dwarf) galaxies   are   
consistent with  the cored energy density profile
for the DM halos  as follows:
\begin{equation}
\rho_d =\frac{\rho_{C}}{1+(r/R_{C})^2},
\end{equation}
where $\rho_{C}$ and $R_C$ are two free core  parameters (see,
e.g., \cite{Blok}).  
The presence of a mild cusp is still admitted by the data.
Such a  profile leads
to the asymptotic constant  rotation velocity
$v_\infty=(4\pi G\rho_{C}R_C^2)^{1/2}$. 
This behaviour is in qualitative agreement  with that for 
the  previously found one-parameter dark halo profile $\rho_h(r/R_{0})$, which 
results in the asymptotic constant velocity $v_{h\infty}
 =(4\pi  G \rho_{0}R_{0}^2/3)^{1/2}$. But now the parameters $\rho_{0}$ and
$R_{0}$ are, in fact, not independent ensuring the fixed
universal $v_{h\infty}=\upsilon_s$.
To disentangle the parameters one should use a three-parameter lacuna profile  
accounting  for  the fracture-halo interference.
The latter  gets significant with the interference
parameter $\de_i={\cal O}(1)$, or with the  scalar
radii  for fracture and halo related as $r_s\sim \upsilon_s R_{0}$, 
$\upsilon_s\ll 1$,
what is not quite
unrealistic. Other factors such as  lacuna asphericity or rotation, 
influence of the scalar-field potential,  influence of the distributed
non-gravitational matter, etc,  may also be of importance.\footnote{In
particular, despite  the fact that  LSB galaxies are 
DM-domi\-nated,  the account for their star disks  may be important for
RCs near the origin.}$^,$\footnote{From the DM
point of view    the dark lacunas may rather   be
considered  as  scalar  ``lumps'' in   asymptotically
non-flat metric. This is to be contrasted with alternative   attempts  at
treating galaxies  in
the framework of GR with a scalar field by means of  scalar 
lumps in  asymptotically flat metrics  (cf., e.g.,~\cite{Wett}).}

With these caveats, the dark lacunas consisting of  a stabilizing
super-massive dark fracture at the origin surrounded by a dark halo might
serve  as a prototype model for  the  galaxy DM frames, to be supplemented
ultimately by the distributed  non-gravitational  matter. 
With  $\ka_g=M_P/\sqrt{8\pi}=2.4\times 10^{18}$
GeV, $M_P=1/\sqrt{G}$ standing for the Planck mass,
the asymptotic rotation velocity in galaxies
$v_{\infty} \sim 10^{-3}$  (in units
of the speed of light)   at its  face value results in 
$\upsilon_s= \sqrt{2}\ka_s/\ka_g\sim 10^{-3}$.
This implies  that the mass  scale appropriate to  the scalar mode,
$\ka_s\sim 10^{15}$~GeV,  is to be of the order of  GUT scale. 
Could it be more than just a coincidence, with a common origin of the two
scales (if any)?

\paragraph{Long-distance gravity modification: vacuum vs.\ Lagran\-gian\ }

In the end  let us present several comments concerning conceivable 
long-distance gravity modification in  the context of galaxy DM. 

(\/{\em i\/}) {\em Spontaneous vacuum modification\ }
The present approach to galaxy DM 
comprises  an explicit modification  of the gravity Lagrangian at a fixed 
ultraviolate mass scale $\kappa_s \sim 10^{-3} \kappa_g$. The effective theory
is to be valid up to  the high scales $\mu \leq \ka_s$.
An infrared  parameter   $\La_0$ 
(or, equivalently, a  long-distance  scale $R_0$)  
appears  at random due to  spontaneous breaking  of a 
global symmetry within a  lacuna. 
The randomness of the infrared parameter 
insures more flexibility of the theory in the context of galaxy~ DM.

(\/{\em ii\/}) {\em Explicit kinetic Lagrangian  modification\ } 
The attempts at the explicit
infrared  Lagrangian modifications  in the context of galaxy DM
are numerous.  First of all, one may mention  the so-called
$f(R)$-gravity,  without  or  with  a pro\-per  modification of matter
Lagrangian (cf., e.g, \cite{fR,fR2}). In this case, having no specific 
degree(s) of freedom,  DM is just mimicked by the modification
of the long-distance tensor gravity.\footnote{ At that,  
a looked-for putative matter   is rather ``missing''  than the  dark
one, being beyond the reach of  direct searches.} 
A related approach is given by the 
scalar-field theories  with a non-canonical 
kinetic term, supplemented ultimately by a repulsive 
potential in Lagran\-gian (cf., e.g., \cite{Bertacca,Gauthier}),
etc. Moreover, one may envisage the case wth a ghost quadratic kinetic
term.\footnote{For the scalar-field ghost condensation as an alternative to DM
in the context of the Universe  cf, e.g.,~\cite{ghost}.}
By construction, so   modified   Lagrangians are given by  some    functions 
$f( R/\mu_I^2)$  or  $f(
\nabla\si\cdot\nabla\si/\mu_I^4)$, etc.,  depending  explicitly on a
fixed infrared  mass scale $\mu_I\ll \ka_g$. 
Ultimately, this would  imply that   a more fundamental theory should settle
down  already on the relatively low scales $\mu\ge \mu_I$.\footnote{It goes
without saying that UBG could a priori
admit, if desired,  an arbitrary infrared modification of the kinetic
Lagrangians both for the  tensor and scalar modes.  Presently such
modifications are left aside.}

\section{~Conclusion and  prospects}
\label{sec:10}

The Unimodular Bimodal Gravity (UBG)  is a
theoretically viable development of UR and GR with a scalar field.
It retains the principle
ingredients of GR -- the general covariance and  masslessness of
the transverse-tensor graviton. 
Beyond GR, the theory  comprises in metric  a  pro\-pagating compression
mode -- the scalar graviton/systolon. 
With the latter treated as DM, UBG presents a unified description of  the
(tensor) gravity and DM. The appearance of a
physically well-motivated scalar field is the crucial point of the theory. In
its reduced version, with the spontaneously broken dynamical global  symmetry, 
the theory, being rather restrictive, is apt to result in
a number of  definite predictions. In particular,  in the static
spherically symmetric case it predicts   peculiar space
structures --  the dark lacunas -- consisting of  a compact singular dark  
fracture (a scalar-dressed BH) at the origin
surrounded by an extended soft-core dark halo. Enjoying 
the property of gravitational confine\-ment,  with the  logarithmic potential
of gravitational attraction  at the periphe\-ry,   the dark lacunas 
ensure   asymptotic flattening of  RCs and may serve  in cosmology  as the
DM frames  for  galaxies.

The   scalar graviton/systolon  physics  pre\-sents
conceivably a perspective field of the future investigations, both theoretical
and phenomenological. In particular, the influence of the (near) massless
scalar field may drastically change the structure of the GR BHs both at
their event horizon and at  asymptotic due to appearance of dark lacunas. 
Further studying the latter ones, as well as dark fractures and
halos, including
their more subtle  aspects such as a  putative asphericity or  rotation, 
influence of  the scalar-graviton  potential and the distributed  
non-gravitatio\-nal matter, application to various  types of galaxies as well
as to galaxy clusters is in order.  The application of the  theory to evolution
of the Universe  is likewise urgent to verify the theory (if any)  and unravel
ultimately  the mystery  of DM and DE.

\paragraph{Acknowledgement}
Thanks are due  to  I.~Yu.\  Polev for   assistance  with numerical 
calculations.

\end{document}